\documentclass{article}

\PassOptionsToPackage{numbers, compress}{natbib}
\usepackage[final,main]{neurips_2026}

\usepackage[utf8]{inputenc}
\usepackage[T1]{fontenc}
\usepackage{url}
\usepackage{booktabs}
\usepackage{microtype}
\usepackage{amsmath,amsfonts,amssymb,amsthm}
\usepackage{graphicx}
\usepackage{float}
\usepackage{algorithm}
\usepackage{algpseudocode}
\usepackage{subfigure}
\usepackage{adjustbox}
\usepackage{bbm}
\usepackage{enumitem}

\usepackage[normalem]{ulem}
\usepackage{hyperref}
\hypersetup{hidelinks}

\newtheorem{theorem}{Theorem}[section]

\theoremstyle{remark}
\newtheorem{definition}[theorem]{Definition}

\allowdisplaybreaks

\graphicspath{{figure/}}

\title{Rank-Transformed Dissimilarity Profiles for High-Dimensional Classification}

\author{Xiangbo Mo \\
  Department of Statistics\\
  University of California, Davis\\
  Davis, CA 95616, USA\\
  \texttt{xbmo@ucdavis.edu}
  \And
  Hao Chen \\
  Department of Statistics\\
  University of California, Davis\\
  Davis, CA 95616, USA\\
  \texttt{hxchen@ucdavis.edu}
}

\begin{document}

\maketitle

\begin{abstract}
Despite advances in representation learning, high-dimensional classification remains challenging in low-sample-size regimes, where the dominant signal may vary across applications and labeled data are often limited. We propose a dissimilarity-profiling classification framework that represents each observation by its class-wise dissimilarity profile, transforming the original feature space into a low-dimensional representation that summarizes how the observation relates to each class. The key idea is to turn an effect of the curse of dimensionality into signal: high-dimensional geometry can induce systematic within-class and between-class dissimilarity patterns under location, scale, or other distributional changes, and these patterns are captured by the class-wise profiles. Building on this representation, we introduce a rank-transformed algorithm that converts dissimilarities into class-wise rank profiles. The proposed method delivers competitive or improved performance relative to commonly used classifiers on two-class, multi-class, network, and real high-dimensional low-sample-size datasets. To clarify the underlying mechanism, we characterize the mean and covariance of a distance-based surrogate, showing how location, scale, and higher-order distributional features enter the profile representation. Contamination experiments show that the rank transformation can improve robustness. The same profile construction applies to non-Euclidean data whenever pairwise dissimilarities can be computed.
\end{abstract}

\section{Introduction}

High-dimensional classification problems arise across many scientific and machine-learning applications. Examples include gene-expression studies with thousands of measured genes for tumor subtype classification~\citep{golub1999molecular, ayyad2019gene}, online review data with high-dimensional 
text-derived features for sentiment classification~\citep{ye2009sentiment,bansal2018sentiment},
and speech signals for classifying speakers' emotions or attitudes~\citep{burkhardt2010database}. In many such settings, the number of features is large relative to the number of labeled observations and the dominant signal separating classes may vary substantially across applications. This makes high-dimensional low-sample-size classification particularly challenging.

A wide range of methods has been developed for high-dimensional classification. Margin-based methods, such as support vector machines (SVM)~\citep{cortes1995support}, remain widely used, with extensions incorporating feature selection or specialized kernels~\citep{ghaddar2018high}. Linear discriminant analysis (LDA) has been adapted to high-dimensional settings through regularization and dimension reduction~\citep{yang2014regularized,li2005robust}. Distance-based methods, including nearest-neighbor classifiers  and their high-dimensional variants, provide another classical approach~\citep{cover1967nearest,liu2006new}. Rank-based methods for HDLSS classification include the rank distribution classifier (RDC) of \mbox{\citet{makinde2020rank}}, which compares an observation's spatial-rank outlyingness across classes. Ensemble approaches, such as random forests~\citep{breiman2001random} and random-projection ensembles~\citep{cannings2017random}, further improve predictive stability through aggregation. More recently, neural network models have achieved strong empirical performance across many domains, including multilayer perceptrons (MLPs) for classification of vector-valued data~\citep{popescu2009multilayer}, convolutional neural networks for image and spatial data~\citep{lecun1989backpropagation,shi2015convolutional,cao2020hyperspectral}, recurrent neural networks for sequential data~\citep{deng2020heart,liu2016recurrent,zhang2021attention}, and generative models for semi-supervised learning~\citep{kingma2014semi,radford2015unsupervised,kim2021unsupervised}.

While numerous classification methods exist, many classical and widely used approaches rely, explicitly or implicitly, on the principle that observations, or their learned/projection-based representations, from the same class tend to be closer than observations from different classes. This principle is often effective in low-dimensional settings, but it can become unreliable in high dimensions due to the curse of dimensionality. 

To illustrate this point, consider a two-class classification problem in
$d=1000$ dimensions with labeled samples
\(X_1,\dots,X_{50} \overset{iid}{\sim} F_X\) and \(Y_1,\dots,Y_{50} \overset{iid}{\sim} F_Y\), 
and a new observation \(W\) to be classified. We generate $X_i = \mathbf{A}x_i$, $Y_j = \mathbf{B}y_j + \mu$, 
where the entries of $x_i$ and $y_j$ are  i.i.d. \(t_5\), \(\mathbf{A}\mathbf{A}^\top=\Sigma\) with \(\Sigma_{r,c}=0.1^{|r-c|}\), \(\mathbf{B}=a\mathbf{A}\), and \(\mu=\mu_0 \mu'/\|\mu'\|\), where $\mu'\sim N_d(0,\mathbf{I}_d)$. Here $\mathbf A$ is the lower-triangular Cholesky factor of
$\Sigma$, with positive diagonal. Performance is evaluated over 50 independent trials, with 50 test samples generated from each class in each trial. By varying \(\mu_0\) and \(a\), the two classes differ in mean and/or covariance.

We evaluate a representative set of commonly used and competitive classifiers, including generalized linear discriminant analysis (GLDA)~\citep{li2005robust}, support vector machines with a radial kernel (SVM)~\citep{cortes1995support}, random forests (RF)~\citep{breiman2001random}, regularized linear models implemented via glmnet~\citep{friedman2010regularization}, gradient-boosting decision trees implemented via LightGBM~\citep{ke2017lightgbm}, a multilayer perceptron (MLP)~\citep{popescu2009multilayer}, the distance-based classifier gMADD~\citep{roy2022generalizations}, the random-projection ensemble classifier (RP-Ensemble)~\citep{cannings2017random}, and the spatial-rank-based rank distribution classifier (RDC)~\citep{makinde2020rank}.

\begingroup
\setlength{\intextsep}{10pt}
\begin{table}[H]
\centering
\footnotesize
\setlength{\tabcolsep}{6pt}
\caption{Misclassification rate (bold: within 0.01 of the lowest)}
\label{introres}
\begin{adjustbox}{max width=\textwidth}
\begin{tabular}{ccccccccccc}
\hline
$\mu_0$ & $a$ & GLDA & SVM & RF & glmnet & LightGBM & MLP & gMADD & RP-Ensemble & RDC \\
\hline
4 & 1 & $\mathbf{0.254}$ & $\mathbf{0.250}$ & 0.309 & 0.416 & 0.435 & 0.332 & 0.466 & 0.383 & 0.399 \\
0 & 1.05 & 0.506 & 0.394 & 0.480 & 0.507 & 0.508 & 0.506 & $\mathbf{0.310}$ & 0.502 & 0.521 \\
\hline
\end{tabular}
\end{adjustbox}
\end{table}
\endgroup

\normalsize

Table~\ref{introres} reports the average misclassification rates. GLDA and SVM perform best when the two classes differ in mean, whereas gMADD performs best  when the two classes differ in variance, with most standard methods close to  chance in that setting. Thus, no single baseline in this comparison adapts well to both types of signal. This motivates a representation that can capture location, scale, and other distributional differences reflected in the geometry of high-dimensional data.

We propose a dissimilarity-profiling framework for high-dimensional classification. The central idea is to represent each observation by its class-wise dissimilarity profile, a low-dimensional summary of how the observation relates to each class through pairwise dissimilarities. This representation turns high-dimensional distance patterns into useful signal: differences in mean, scale, or other distributional features can induce systematic patterns among within-class and between-class dissimilarities, and these patterns are captured by the class-wise profiles. Building on this framework, we introduce a rank-transformed dissimilarity-profiling classification algorithm. The method converts pairwise dissimilarities into class-wise rank profiles, producing a low-dimensional representation whose dimension equals the number of classes. Whereas RDC scores each class using spatial-rank outlyingness, our method classifies observations using the joint pattern across classes. The rank transformation improves robustness to extreme distances, and the resulting profiles can be classified using standard low-dimensional methods such as quadratic discriminant analysis. The profile construction uses only pairwise dissimilarities, so it applies to non-Euclidean data.

The remainder of the paper is organized as follows. Section~\ref{sec:method} examines the motivating example in detail and introduces the proposed dissimilarity-profiling framework along with the rank-transformed classification algorithm. Section~\ref{sec:simulation} evaluates the performance of the method across two-class, multi-class, network, and real-data settings. To provide deeper insight into the approach, Section~\ref{sec:analysis} investigates key quantities and mechanisms underlying its effectiveness through a distance-based surrogate. Section~\ref{sec:conclusion} concludes with a discussion of limitations and broader impact.

\section{Dissimilarity-profiling classification}
\label{sec:method}

\subsection{Intuition}

\begin{figure}[!htbp]
\centering
\includegraphics[width=0.245\textwidth]{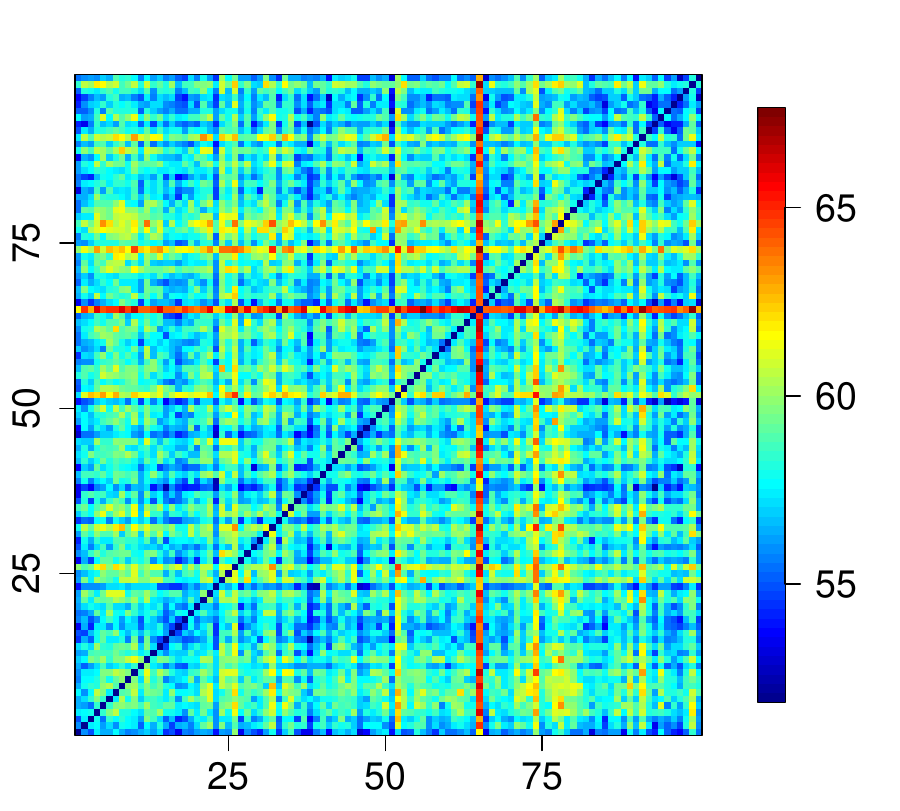}
\includegraphics[width=0.245\textwidth]{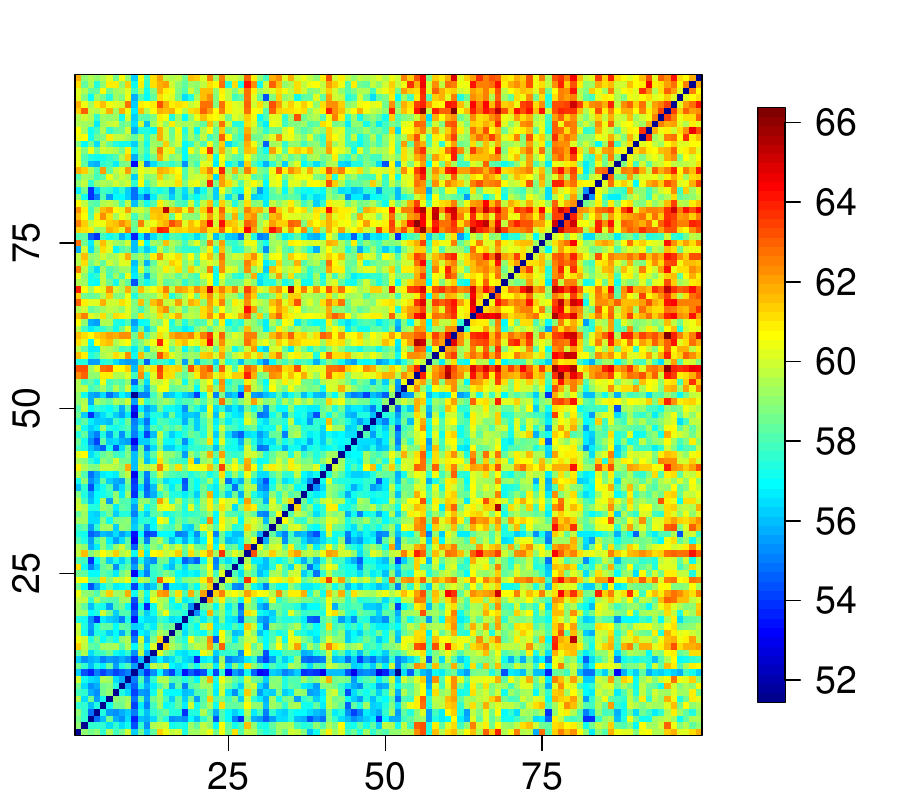}
\includegraphics[width=0.245\textwidth]{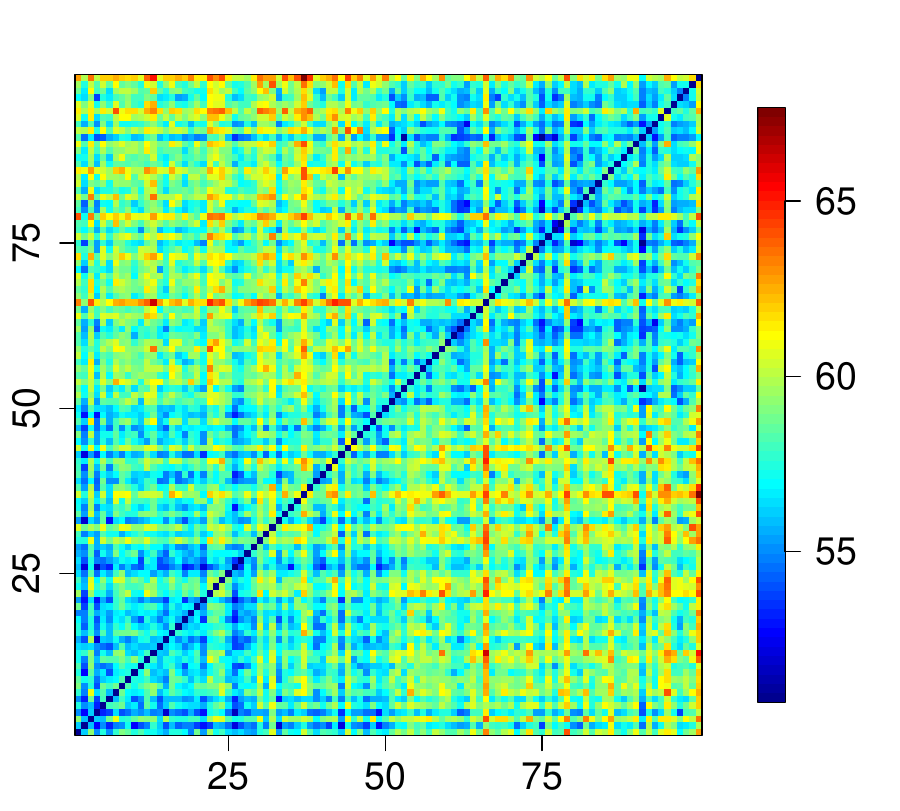}
\includegraphics[width=0.245\textwidth]{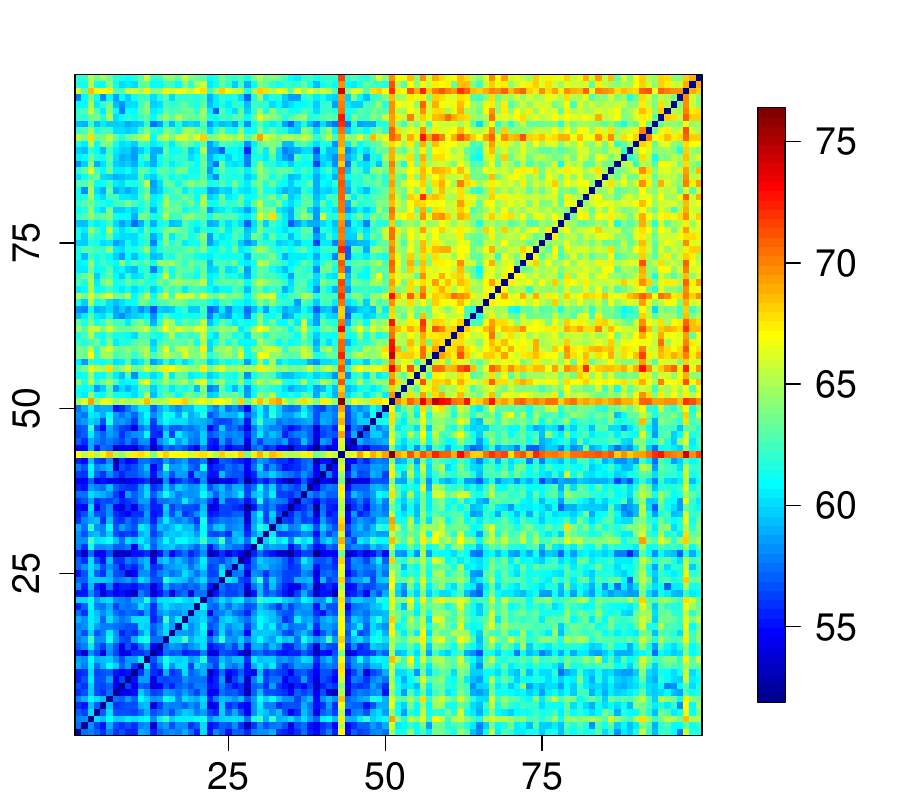}
\caption{Heatmaps of pairwise distances under mean- and variance-difference settings. 
Observations are ordered as $X_1,\ldots,X_{50},Y_1,\ldots,Y_{50}$. 
From left to right, the panels correspond to: mean difference with $(\mu_0,a)=(4,1)$, 
variance difference with $(\mu_0,a)=(0,1.05)$, stronger mean difference with 
$(\mu_0,a)=(15,1)$, and stronger variance difference with $(\mu_0,a)=(0,1.15)$. 
 }
\label{fig:heatmap}
\end{figure}

The proposed approach is motivated by the observation that distributional differences between classes can be reflected in the pattern of within-class and between-class inter-point dissimilarities. We first illustrate this idea in the two-class setting from Table~\ref{introres}. Let $D_{XX}$ denote the distance between two observations from class $X$, $D_{YY}$ the distance between two observations from class $Y$, and $D_{XY}$ the distance between one observation from each class. Figure~\ref{fig:heatmap} shows heatmaps of pairwise distances under the mean- and variance-difference settings from Table~\ref{introres}, together with stronger-signal versions that make the corresponding distance patterns more visible.

Under a mean difference, the between-class distance $D_{XY}$ tends to be larger than 
the within-class distances $D_{XX}$ and $D_{YY}$, as shown in the first and third 
panels of Figure~\ref{fig:heatmap}. In contrast, under a variance difference where 
class $Y$ is more dispersed than class $X$, the distances often follow the ordering $D_{XX} < D_{XY} < D_{YY}$, as shown in the second and fourth panels. This phenomenon reflects the geometry of high-dimensional data: observations from a more dispersed distribution can be farther apart from one another than from observations in a more concentrated distribution.
Similar high-dimensional inter-point distance behavior 
has been noted in the two-sample testing literature~\citep{chen2017new}. Here, we use this behavior as signal for classification through class-wise 
dissimilarity profiles. The key idea is to summarize how each observation relates 
to each class through within-class and between-class dissimilarities. In the 
two-class setting, an observation from class $X$ can be described by its 
dissimilarities to classes $X$ and $Y$, involving $(D_{XX},D_{XY})$, while an 
observation from class $Y$ can be described by $(D_{XY},D_{YY})$.

The same idea extends naturally to multi-class classification. For a $k$-class 
problem with class labels $1,\ldots,k$, let $D_{ij}$ denote the dissimilarity 
between an observation from class $i$ and an observation from class $j$. For an 
observation from class $i$, we summarize its relation to all classes by the vector
$D_i = (D_{i1},D_{i2},\ldots,D_{ik}).$
If class $i$ differs from class $j$, this difference may be reflected in the 
joint behavior of $D_{ii}$, $D_{ij}$, and $D_{jj}$. Therefore, the class-wise 
dissimilarity profile provides a compact representation that can capture different 
forms of distributional separation.
\subsection{Proposed method}

  For a positive integer $r$, write $[r]=\{1,\ldots,r\}$. 
Let $S=\{(Z_i,g_i): i\in[N]\}$ be a training set, where $Z_i$ is the $i$-th observation and $g_i\in\{1,\dots,k\}$ its class label. The observations are assumed independent, with $Z_i \sim F_{g_i}$. Let $n_j=\sum_{i=1}^N \mathbbm{1}\{g_i=j\}$ denote the number of observations in class $j$. Given a new observation $W$, drawn from one of the distributions $\{F_j\}_{j=1}^k$, the goal is to assign $W$ to its class.
The proposed classification framework consists of the following general steps:
\begin{enumerate}\item construct a pairwise dissimilarity matrix among the training observations;
    \item summarize each observation by its class-wise dissimilarity profile;
    \item compute the corresponding class-wise dissimilarity profile for a new observation;
    \item classify the new observation in the resulting low-dimensional profile space.
\end{enumerate}

    This framework is flexible. The dissimilarity measure can be chosen according to  the data type and the scientific question. For vector-valued data, Euclidean distance is a natural choice, while alternatives such as Manhattan, Chebyshev, or Minkowski distances may be preferable in settings with outliers or coordinate-wise  heterogeneity. For structured data, the dissimilarity should reflect the underlying object type, such as Frobenius distance for matrices or Hamming distance for graphs or discrete data. Similarly, the class-wise summary can be chosen according to the problem: the mean provides a simple default, while medians, trimmed means, or Winsorized means may improve robustness. After this transformation, classification is performed in a $k$-dimensional profile space, where standard classifiers can be applied when the number of classes is moderate.

    In this paper, we focus on a rank-transformed implementation of this framework. Starting from pairwise squared Euclidean distances, we convert distances into ranks and then compute class-wise average ranks as the profile representation. Since squaring preserves the ordering of nonnegative distances, using Euclidean distances gives the same rank profiles. The rank transformation reduces sensitivity to extreme distances while preserving the relative ordering information that is central to the proposed approach. We then classify the resulting profiles using quadratic discriminant analysis. The full procedure is summarized in Algorithm~\ref{rank_algorithm}.

\begin{algorithm}[!htbp]
\caption{Rank-Transformed Dissimilarity-Profiling Classification Algorithm}
\label{rank_algorithm}

\noindent Input: training vectors $Z_i\in\mathbb{R}^d$ and their labels, and a test vector $W\in\mathbb{R}^d$.
\begin{itemize}[leftmargin=1.2em]
    \item [1.] Compute the pairwise dissimilarity matrix $D \in \mathbb{R}^{N \times N}$, where $D[i,j] = \Vert Z_i - Z_j \Vert_2^2 $, and the rank matrix $R \in \mathbb{R}^{N \times N}$, where $ R[i,j] $ is the average rank of $\ D[i,j] \ $ in the  $ j^{th}$  column.
    \item [2.] Compute the rank mean matrix $M^{(R)} \in \mathbb{R}^{N \times k}$: $M^{(R)}[i,j]= \frac{1}{n_j - \mathbbm{1}_{g_i = j}} \sum\limits_{{\scriptscriptstyle  l: g_l = j, l \neq i}} R[i,l].$
    \vspace{-2mm}
    \item [3.] Let $v_i=M^{(R)}[i,\cdot]^\top$. For a test observation $W$:
    \begin{itemize}[leftmargin=1em]
        \item [•] Compute the mid-empirical rank score of each
        test dissimilarity against the corresponding training column:
     $R_{W}[i] = \frac{1}{2}+\sum_{t=1}^N \mathbbm{1}_{\Vert Z_t- Z_i \Vert_2^2 < \Vert W- Z_i \Vert_2^2} + \frac{1}{2}\sum_{t=1}^N \mathbbm{1}_{\Vert Z_t- Z_i \Vert_2^2 = \Vert W- Z_i \Vert_2^2}.$
        \item [•] 
        $M^{(R)}_W[j] = \frac{1}{n_j}\sum_{\ell:g_\ell=j} R_W[\ell],\quad j\in[k].$
        \item [•] $ g(W) = \operatorname*{arg\,max}\limits_{{\scriptstyle j}} \left \{ -\frac{1}{2} \log \vert \hat{\Sigma}^{(R)}_j \vert - \frac{1}{2}(M^{(R)}_W-\hat{\mu}^{(R)}_j)^{\top} (\hat{\Sigma}^{(R)}_j)^{-1} (M^{(R)}_W-\hat{\mu}^{(R)}_j) + \log \frac{n_j }{N} \right \}, $
    $  \hat{\mu}^{(R)}_j = \frac{1}{n_j} \sum\limits_{{\scriptscriptstyle i:g_i =j}} v_i,$
    $\hat{\Sigma}^{(R)}_j = \frac{1}{n_j-1} \sum\limits_{{\scriptscriptstyle i:g_i =j}} \left( v_i- \hat{\mu}^{(R)}_j \right)\left( v_i- \hat{\mu}^{(R)}_j \right)^\top.$
    \vspace{-2mm}
    \end{itemize}

\end{itemize}

\end{algorithm}

Algorithm~\ref{rank_algorithm} uses unregularized QDA when
the empirical covariance matrix of each class in profile space is positive
definite. This requires $n_j>k$ for every class, although the sample-size
condition alone does not guarantee invertibility. If an empirical covariance
matrix is singular, the same profiles can be used with regularized QDA or
another classifier.

    We applied Algorithm~\ref{rank_algorithm} to the same two settings used in Table~\ref{introres}. The results are reported in Table \ref{tab:intro-new}. Under mean differences, the proposed method is competitive with leading methods such as SVM and GLDA. Under variance differences, it achieves the lowest misclassification rate among the methods considered. These results show that the rank-transformed dissimilarity profiles can adapt to both location and scale differences in this motivating example.

\begin{table}[H]
\centering
\footnotesize
\setlength{\tabcolsep}{4pt}
\caption{Misclassification rate under the same settings as Table~\ref{introres} (bold: within 0.01 of the lowest).}
\label{tab:intro-new}
\begin{adjustbox}{max width=\textwidth}
\begin{tabular}{cccccccccccc}
\hline
$\mu_0$ & $a$ & Rank profile & GLDA & SVM & RF & glmnet & LightGBM & MLP & gMADD & RP-Ensemble & RDC \\
\hline
4 & 1 & 0.278 & $\mathbf{0.254}$ & $\mathbf{0.250}$ & 0.309 & 0.416 & 0.435 & 0.332 & 0.466 & 0.383 & 0.399 \\
0 & 1.05 & $\mathbf{0.266}$ & 0.506 & 0.394 & 0.480 & 0.507 & 0.508 & 0.506 & 0.310 & 0.502 & 0.521 \\
\hline
\end{tabular}
\end{adjustbox}
\end{table}

\begin{figure}[!t]
\centering
\includegraphics[width=0.245\textwidth]{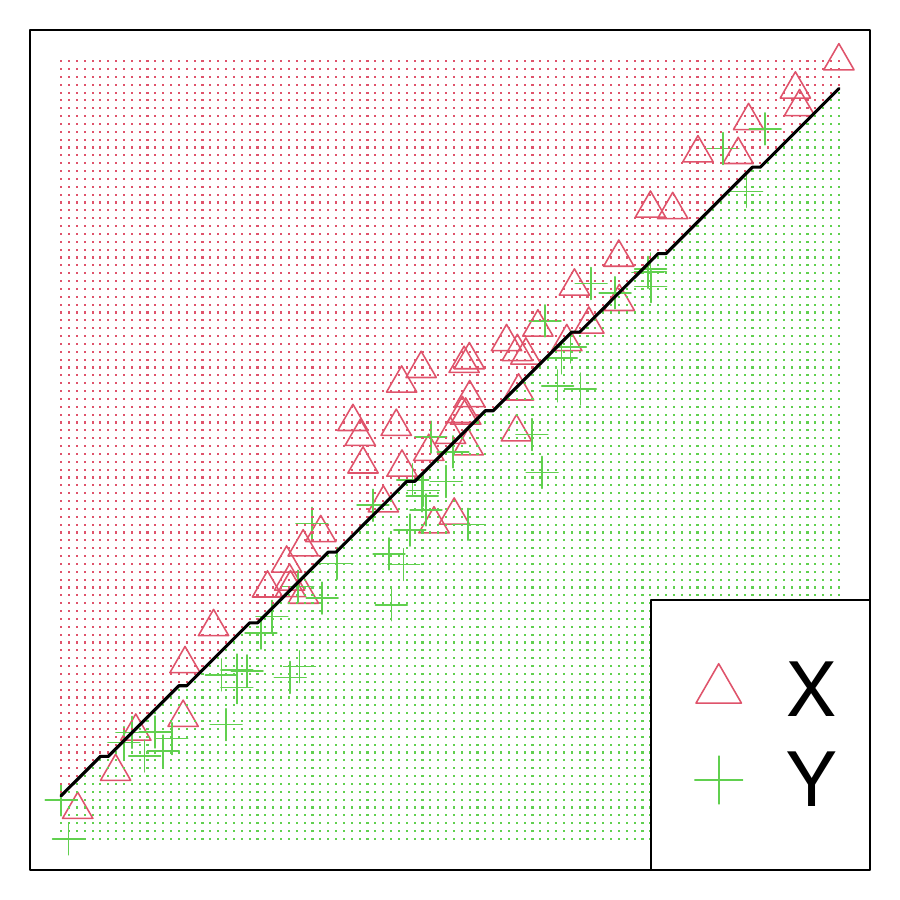}
\includegraphics[width=0.245\textwidth]{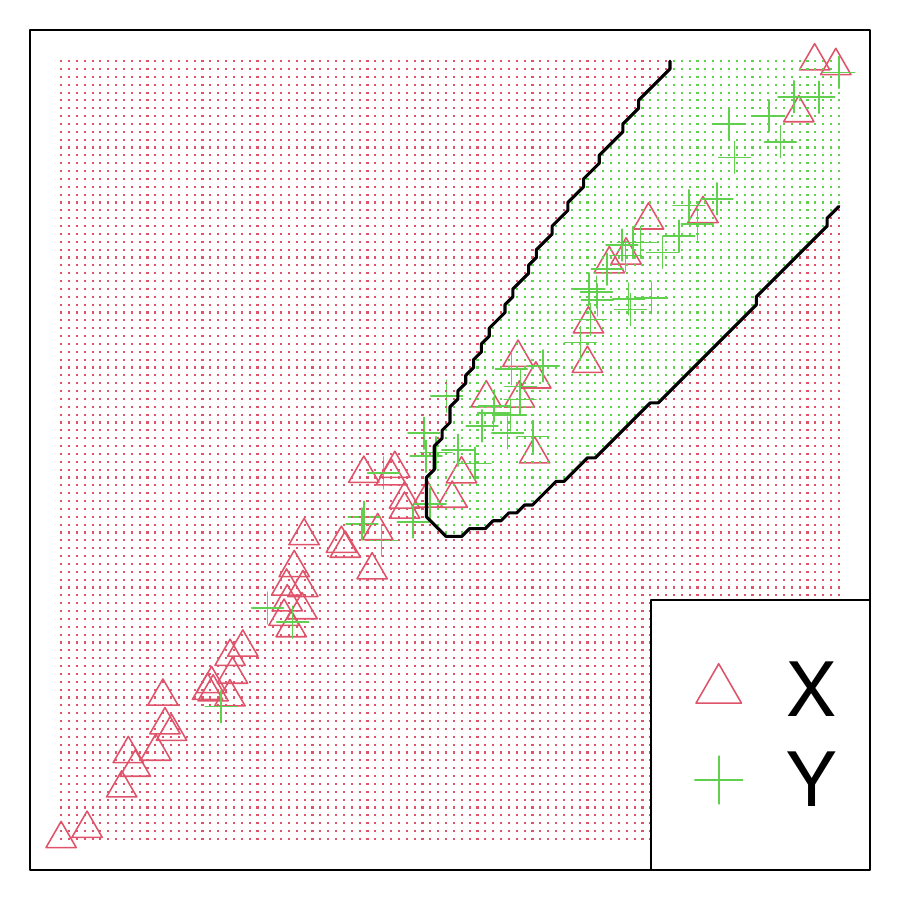}
\includegraphics[width=0.245\textwidth]{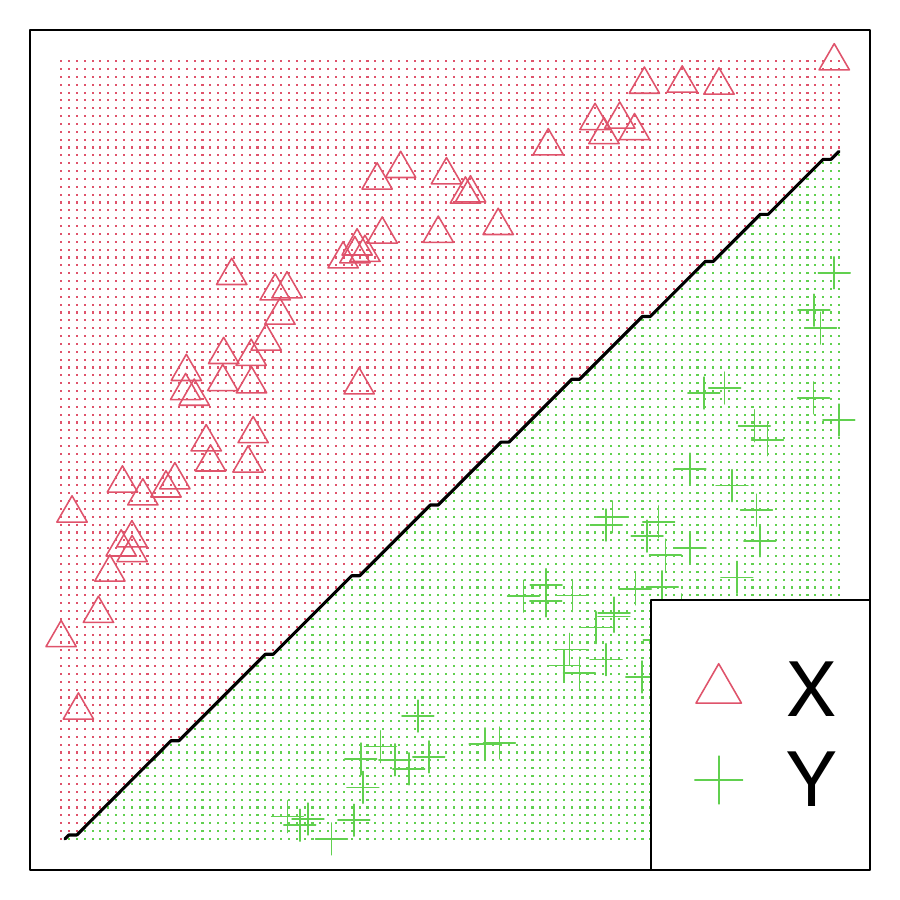}
\includegraphics[width=0.245\textwidth]{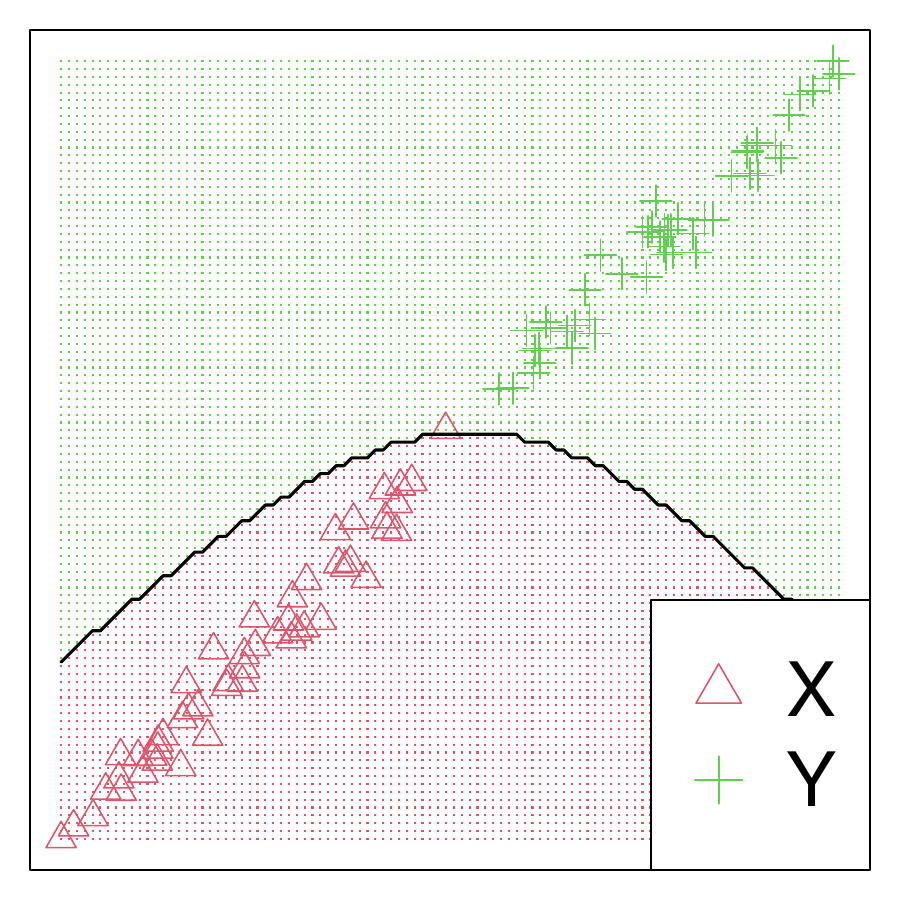}
\caption{Class-wise rank profiles $M^{(R)}[i,\cdot]$ for the training observations. 
From left to right: $(\mu_0,a)=(4,1)$, $(0,1.05)$, $(15,1)$, and $(0,1.15)$. 
Points are colored by class, and the fitted QDA decision boundaries are shown. }
\label{fig:decision_dist}
\end{figure}

   To visualize the effect of the profile transformation, Figure~\ref{fig:decision_dist} plots the class-wise rank profiles $M^{(R)}[i,\cdot]$ for the training observations. The two classes become separated in the profile space under both mean- and variance-difference settings. This illustrates how the first two steps of the algorithm transform high-dimensional pairwise dissimilarities into a low-dimensional representation that captures the relevant class-separating structure.

\section{Performance comparisons}
\label{sec:simulation}
 \vspace{-0.5em}
We further evaluate the proposed rank-transformed dissimilarity-profiling method 
across two-class, multi-class, network, and real-data settings. Unless otherwise 
specified, we compare with the same baseline methods used in the motivating examples. 
The numerical tables report 50 repetitions per setting.
Gene-expression experiments use 50 QDA-admissible stratified splits selected
by the training-only rule in Section~\ref{sec:nutt_counts}.
Appendix~\ref{app:sd} reports standard deviations; Monte Carlo standard
errors are identified separately.

 \vspace{-0.5em}
\subsection{Two-class classification}
\label{sec:two_class}
 \vspace{-0.4em}

We first consider two-class classification for
$d=1000$ vector-valued data, using 50 training and 50 independent test
observations per class. For S1--S4, we generate
$X=\mathbf A x$ and $Y=a\mathbf A y+\mu$, where the coordinates of $x$ and
$y$ are i.i.d. from $F_x$ and $F_y$, respectively. Here $\mathbf A$ is the
lower-triangular Cholesky factor of $\Sigma$, with
$\Sigma_{r,c}=0.1^{|r-c|}$, and $\mu=\mu_0 u/\|u\|$ for
$u\sim N_d(0,I_d)$. Thus, $\mu_0$ controls the mean separation and $a$ the
relative scale; their values are given in Table~\ref{1res}.
We consider five scenarios:

\begin{itemize}
    \item Scenario S1: $F_x=N(0,1)$ and $F_y=N(0,1)$;
    \item Scenario S2: $F_x=t_5$ and $F_y=t_5$;
    \item Scenario S3: $F_x=\chi^2_5-5$ and $F_y=\chi^2_5-5$;
    \item Scenario S4: $F_x=N(0,1)$ and $F_y=t_5$;
    \item Scenario S5: $F_X=N_d(0,I_d)$ and $F_Y=N_d(0,\Sigma_{0.9})$, where $(\Sigma_{0.9})_{ij}=0.9^{|i-j|}$.
\end{itemize}

Table~\ref{1res} reports the misclassification rates for S1--S5. The proposed method achieves the lowest error in all three S1 settings, most settings involving scale differences, and the correlation setting S5, and has near-zero error for Gaussian versus $t_5$. SVM performs best for the non-Gaussian location-shift alternatives, while gMADD performs best for the $t_5$ scale alternative. Appendix~\ref{app:covariance_pilots} reports additional experiments on correlation strength and signed-block covariance differences.

\begin{table}[H]
\centering
\footnotesize
\setlength{\tabcolsep}{3pt}
\caption{Two-class misclassification rate for $1{,}000$-dimensional data (bold: within 0.01 of the lowest). Dashes indicate that $\mu_0$ and $a$ do not apply to the S5 parameterization.}
\label{1res}
\begin{adjustbox}{max width=\textwidth}
\begin{tabular}{ccccccccccccc}
\hline
 & $\mu_0$ & $a$ & Rank profile & GLDA & SVM & RF & glmnet & LightGBM & MLP & gMADD & RP-Ensemble & RDC \\
\hline
S1 & 6 & 1 & $\mathbf{0.023}$ & $\mathbf{0.025}$ & $\mathbf{0.024}$ & 0.114 & 0.200 & 0.285 & 0.170 & 0.299 & 0.171 & 0.213 \\
S1 & 0 & 1.1 & $\mathbf{0.019}$ & 0.495 & 0.129 & 0.421 & 0.503 & 0.496 & 0.486 & 0.047 & 0.506 & 0.591 \\
S1 & 6 & 1.1 & $\mathbf{0.004}$ & 0.038 & $\mathbf{0.005}$ & 0.110 & 0.212 & 0.299 & 0.171 & 0.033 & 0.189 & 0.586 \\
S2 & 6 & 1 & 0.109 & $\mathbf{0.103}$ & $\mathbf{0.095}$ & 0.159 & 0.291 & 0.327 & 0.210 & 0.360 & 0.271 & 0.272 \\
S2 & 0 & 1.1 & 0.118 & 0.501 & 0.189 & 0.445 & 0.503 & 0.508 & 0.501 & $\mathbf{0.095}$ & 0.498 & 0.531 \\
S2 & 6 & 1.1 & $\mathbf{0.044}$ & 0.127 & $\mathbf{0.048}$ & 0.156 & 0.315 & 0.326 & 0.254 & 0.067 & 0.279 & 0.522 \\
S3 & 6 & 1 & 0.428 & $\mathbf{0.401}$ & $\mathbf{0.396}$ & 0.412 & 0.483 & 0.476 & 0.444 & 0.498 & 0.449 & 0.458 \\
S3 & 0 & 1.1 & $\mathbf{0.084}$ & 0.501 & 0.170 & 0.371 & 0.494 & 0.491 & 0.497 & 0.109 & 0.503 & 0.537 \\
S3 & 6 & 1.1 & $\mathbf{0.084}$ & 0.405 & 0.128 & 0.306 & 0.492 & 0.485 & 0.452 & 0.103 & 0.457 & 0.553 \\
S4 & 0 & 1 & $\mathbf{0.000}$ & 0.468 & $\mathbf{0.000}$ & 0.401 & 0.498 & 0.500 & 0.468 & $\mathbf{0.007}$ & 0.488 & 0.520 \\
S5 & -- & -- & $\mathbf{0.191}$ & 0.537 & 0.304 & 0.397 & 0.507 & 0.493 & 0.429 & 0.377 & 0.510 & 0.645 \\
\hline
\end{tabular}
\end{adjustbox}
\end{table}

We also tested whether supervised feature screening helps
when only four of 100 coordinates carry signal. In the sparse-mean setting,
screening by the Welch statistic using only the training data reduced the
mean test error from $0.247$ in the full space to $0.169$, close to $0.164$
when the four true signal coordinates were supplied. In the equal-trace
variance swap, screening by a combined mean and variance score reduced
the error from $0.496$ to $0.280$, matching the result on the true signal
coordinates. The classifier was the same throughout; only the coordinates
used to compute dissimilarities changed. Appendix~\ref{app:supervised_geometry}
gives the full results and screening rules.

\subsection{Multi-class classification}
 \vspace{-0.5em}

We next consider four-class $d=1000$ data with 50 training and
50 independent test observations per class. For class $i\in[4]$, we generate
$Z_{ij}=a_i\mathbf A z_{ij}+\mu_i$, where the coordinates of $z_{ij}$ are
i.i.d. from $F_z$ and $\mathbf A$ is the same Cholesky factor as in S1--S4.
We set $a_1=a_3=1$, $a_2=a_4=1.1$, $\mu_1=\mu_2=0$, and
$\mu_3=\mu_4=\mu$, with $\mu=12u/\|u\|$ for $u\sim N_d(0,I_d)$.
Thus, the four classes combine two means and two scales. We use:

\hspace{5mm} \raisebox{0.2ex}{\scriptsize$\bullet$}\ Scenario S6: $F_z=N(0,1)$; \hspace{5mm}
\raisebox{0.2ex}{\scriptsize$\bullet$}\ Scenario S7: $F_z=t_5$; \hspace{5mm}
\raisebox{0.2ex}{\scriptsize$\bullet$}\ Scenario S8: $F_z=\chi_5^2-5$.

 RP-Ensemble is omitted because it is formulated for
two-class classification. Table~\ref{tab:multiclass} reports the
multi-class misclassification rates. The proposed method achieves the lowest
error rate in S6 and S8, and is the second-best method in S7, where gMADD
performs best. These results suggest that our rank-profile representation
extends naturally beyond two-class classification and remains effective in
multi-class HDLSS settings involving both mean and scale differences.

\begin{table}[H]
\centering
\footnotesize
\setlength{\tabcolsep}{4pt}
\caption{Multi-class misclassification rate for $1{,}000$-dimensional data.}
\label{tab:multiclass}
\begin{adjustbox}{max width=\textwidth}
\begin{tabular}{cccccccccc}
\hline
 & Rank profile & GLDA & SVM & RF & glmnet & LightGBM & MLP & gMADD & RDC \\
\hline
S6 & $\mathbf{0.023}$ & 0.490 & 0.128 & 0.453 & 0.517 & 0.507 & 0.481 & 0.053 & 0.642 \\
S7 & 0.139 & 0.496 & 0.197 & 0.459 & 0.534 & 0.510 & 0.499 & $\mathbf{0.089}$ & 0.563 \\
S8 & $\mathbf{0.227}$ & 0.581 & 0.310 & 0.498 & 0.699 & 0.631 & 0.626 & 0.492 & 0.694 \\
\hline
\end{tabular}
\end{adjustbox}
\end{table}

\subsection{Network data classification}
We next evaluate the proposed framework on network-valued observations. Random graphs are 
generated from the configuration model $G(v,\vec{k})$, where $v=40$ is the number of vertices and 
$\vec{k}=(k_1,\ldots,k_v)$ is the degree sequence.
The implementation uses the configuration-model generator
in \texttt{igraph}, which permits self-loops and multiple edges. In each
adjacency matrix $A$, $A_{ij}$ is the number of edges between vertices $i$
and $j$, with $A_{ii}$ giving the number of self-loops at vertex $i$.
In each trial, we generate independent training 
graphs $X_i\stackrel{i.i.d.}{\sim}G(v,\vec{k}^x)$, $i\in[30]$, and 
$Y_j\stackrel{i.i.d.}{\sim}G(v,\vec{k}^y)$, $j\in[30]$. For testing, we generate 
$W_{x,i}\stackrel{i.i.d.}{\sim}G(v,\vec{k}^x)$, $i\in[20]$, and 
$W_{y,j}\stackrel{i.i.d.}{\sim}G(v,\vec{k}^y)$, $j\in[20]$.
We consider the following scenarios:
\begin{itemize}[itemsep=1pt, topsep=2pt, parsep=0pt, leftmargin=9mm]
    \item Scenario S9: $k_i^x=10$ for $i=1,\dots,40$; $k_i^y = 10\,\mathbbm{1}\{i \le 40-a\} + 8\,\mathbbm{1}\{i > 40-a\}$.
    \item Scenario S10: $k_i^x=30$ for $i=1,\dots,40$; $k_i^y = 30\,\mathbbm{1}\{i \le 40-a\} + 20\,\mathbbm{1}\{i > 40-a\}$.
\end{itemize}

\begin{table}[!b]
\centering
\footnotesize
\setlength{\tabcolsep}{6pt}
\caption{Misclassification rates on simulated configuration-model graphs.}
\label{network_res_hamming1}
\begin{adjustbox}{max width=\textwidth}
\begin{tabular}{lcccccc}
\hline
 & S9 ($a=4$) & S9 ($a=5$) & S9 ($a=6$) & S10 ($a=4$) & S10 ($a=5$) & S10 ($a=6$) \\
\hline
Rank profile & $\mathbf{0.278}$ & $\mathbf{0.212}$ & $\mathbf{0.157}$ & $\mathbf{0.151}$ & $\mathbf{0.104}$ & $\mathbf{0.066}$ \\
SVM & 0.458 & 0.443 & 0.449 & 0.207 & 0.159 & 0.116 \\
gMADD & 0.373 & 0.291 & 0.252 & 0.268 & 0.190 & 0.152 \\
\hline
\end{tabular}
\end{adjustbox}
\end{table}

For these graphs, we use the edge-count $L_1$ dissimilarity
$d_{\mathrm{net}}(G,\tilde G)=\frac{1}{v(v-1)}
\sum_{i\le j}|A_{ij}-\tilde A_{ij}|$.
Table~\ref{network_res_hamming1} compares our rank-profile classifier,
network SVM, and gMADD using the same graph dissimilarity
$d_{\mathrm{net}}$. Standard vector-based classifiers
would require a separate feature representation of the graphs and are
therefore not included in this comparison.
Our rank-profile classifier achieves the lowest
test error in every network setting.

\subsection{Real data}

We evaluate the proposed method on real high-dimensional low-sample-size datasets.

\subsubsection{Gene-expression data}\label{sec:nutt_counts}

We consider two benchmark gene-expression datasets: the Shipp et al.~\citep{shipp2002diffuse} 
dataset for diffuse large B-cell lymphoma and the Nutt et al.~\citep{nutt2003gene} dataset for 
malignant glioma classification. We use Shipp-2002-v1 and
Nutt-2003-v2. We evaluated all methods on the same 50 distinct
class-stratified splits, using one-third of each class (rounded down) for training and the rest
for testing. Candidate splits were generated in seed order and screened using
only the training data to verify that unregularized QDA could be fitted to both
the rank and distance profiles. The reported errors are conditional on this
fitting requirement.
Table~\ref{tab:realdata} shows that the rank profile has the lowest mean error on Nutt and is within 0.01 of
the best method on Shipp under this split protocol.

\begin{table}[H]
\centering
\footnotesize
\setlength{\tabcolsep}{6pt}
\caption{Misclassification rates on selected real datasets (bold: within 0.01 of the lowest).}
\label{tab:realdata}
\begin{adjustbox}{max width=\textwidth}
\begin{tabular}{cccccccccc}
\hline
 & Rank profile & GLDA & SVM & RF & glmnet & LightGBM & MLP & gMADD & RP-Ensemble \\
\hline
Shipp & $\mathbf{0.178}$ & $\mathbf{0.174}$ & 0.250 & 0.222 & 0.238 & 0.250 & $\mathbf{0.168}$ & 0.316 & 0.201 \\
Nutt & $\mathbf{0.168}$ & 0.201 & 0.464 & 0.249 & 0.407 & 0.500 & 0.550 & 0.369 & 0.197 \\
\hline
\end{tabular}
\end{adjustbox}
\end{table}

\begingroup
\subsubsection{PROTEINS graph data}

We further evaluate the method on PROTEINS~\citep{borgwardt2005protein}, a collection of 1,113 protein graphs distributed
through TUDataset~\citep{morris2020tudataset}. The graphs have no
natural vertex correspondence, so we compare their local structure using
concatenated Weisfeiler--Lehman subtree counts at iterations $h=0,1,2$~\citep{shervashidze2011weisfeiler}, normalize each vector to unit
Euclidean norm, and use Algorithm~\ref{rank_algorithm} with squared
Euclidean dissimilarity on this permutation-invariant graph representation.
We report balanced test error for all methods on the same 50
stratified splits, using one-third of each class (rounded down) for training.
The WL SVM is a
linear SVM on the normalized WL2 features. For GLDA, radial SVM, RF, glmnet,
LightGBM, and RP-Ensemble, coordinates with training-set variance at most
$10^{-12}$ are removed; the same retained-coordinate mask is applied to the
test data.
Table~\ref{tab:proteins} shows that the rank profile is within 0.01 of random forest,
the lowest-error method, and has lower mean error than the remaining methods.

\begin{table}[H]
\centering
\footnotesize
\setlength{\tabcolsep}{2pt}
\caption{Mean balanced errors on the PROTEINS graph dataset (bold: within 0.01 of the lowest).}
\label{tab:proteins}
\begin{adjustbox}{max width=\textwidth}
\begin{tabular}{lccccccccc}
\hline
 & RF & \shortstack{Rank\\profile} & \shortstack{RP-\\Ensemble} & WL SVM & \shortstack{gMADD-style\\WL2} & LightGBM & GLDA & glmnet & \shortstack{Radial\\SVM} \\
\hline
Mean balanced error & $\mathbf{0.277}$ & $\mathbf{0.284}$ & 0.291 & 0.301 & 0.302 & 0.310 & 0.316 & 0.331 & 0.449 \\
\hline
\end{tabular}
\end{adjustbox}
\end{table}

\endgroup

\section{Analysis of key quantities}
\label{sec:analysis}

   In this section, we analyze key quantities underlying the proposed framework.
Because the rank transformation introduces nonlinear dependence that is difficult
to characterize directly, we focus on a distance-based surrogate that preserves
the same class-wise profiling structure while replacing ranks by squared Euclidean
distances. This surrogate provides insight into why dissimilarity profiles can
capture different forms of distributional separation. We consider the following two-class setting:
\[
(\star)\quad \begin{aligned}
X_i &= \mathbf A (x_{i1},\ldots,x_{id})^\top +\mu_X,\ 
\mathbf A\in\mathbb R^{d\times d},\ 
x_{ik}\stackrel{i.i.d.}{\sim}F_x,\ 
\mathbb E x_{11}=0,\  \mathbb E x_{11}^4<\infty,\ i\in[n],\\
Y_j &= \mathbf B (y_{j1},\ldots,y_{jd})^\top +\mu_Y,\  
\mathbf B\in\mathbb R^{d\times d},\  
y_{jk}\stackrel{i.i.d.}{\sim}F_y,\ 
\mathbb E y_{11}=0,\  \mathbb E y_{11}^4<\infty,\ j\in[m].
\end{aligned}
\]
The two samples are independent, and $F_x,F_y$ do not vary
with $d$. We assume $n,m\geq2$.
Every test observation considered below is independent of the training
samples and has the corresponding class distribution.
This setting allows the two classes to differ through their means, covariance 
structures, and marginal distributions. Our goal is to use this setting to 
understand how class-wise dissimilarity profiles reflect different types of 
distributional separation.

\subsection{Distance-based surrogate}

For the theoretical analysis, we use the distance-based surrogate
in Algorithm~\ref{dist_algorithm} (Appendix~\ref{app_dist}). It follows
Algorithm~\ref{rank_algorithm} but averages squared Euclidean distances rather
than ranks within each class, then applies QDA to the resulting profiles.
In the absence of outliers, the two algorithms perform comparably across the
settings in Tables~\ref{tab:dr1} and~\ref{tab:dr2}.

Table~\ref{outres} compares the two algorithms under several
forms of training contamination. In OA--OC, replacing three training
observations with shifted and rescaled observations, heavy-tailed
observations, or observations with single-coordinate spikes increased the
surrogate's mean test error from $14.5\%$ to $34.9\%$--$50.0\%$, while the
rank-based classifier's error went from $14.9\%$ to $15.0\%$--$16.9\%$.
In the separate OD setting, replacing seven training observations with
outliers increased the surrogate's error from $2.3\%$ to $27.0\%$, compared
with an increase from $2.4\%$ to $6.9\%$ for the rank-based classifier.
These results motivate using Algorithm~\ref{rank_algorithm} in practice.

\subsection{Mean and covariance of distance profiles}

We first examine the mean and covariance structure of the distance profiles under 
Setting~$(\star)$. Define
\[
\begin{aligned}
D(X_i)
&=\biggl(
\frac{1}{n-1}\sum_{\ell\neq i}\|X_\ell-X_i\|_2^2,\,
\frac{1}{m}\sum_{\ell=1}^m\|Y_\ell-X_i\|_2^2
\biggr)^\top,\\
D(Y_j)
&=\biggl(
\frac{1}{n}\sum_{\ell=1}^n\|X_\ell-Y_j\|_2^2,\,
\frac{1}{m-1}\sum_{\ell\neq j}\|Y_\ell-Y_j\|_2^2
\biggr)^\top .
\end{aligned}
\]
The following theorem gives the expectations and covariance matrices of these 
distance profiles.

\begin{theorem}

\label{theorem1}

 Under Setting ($\star$), we have the following, with $\Sigma_{D_X},\Sigma_{D_Y}$ defined in Definition \ref{def2}:
\begin{align*}
    \mathbb{E} (D(X_i))  
   = &\left(
    \begin{matrix}
    2 \Vert \mathbf{A} \Vert_F^2 \sigma^2(F_x) \\
    \Vert \mathbf{A} \Vert_F^2 \sigma^2(F_x) + \Vert \mathbf{B} \Vert_F^2 \sigma^2(F_y) + \Vert \mu_Y - \mu_X \Vert_2^2
    \end{matrix} \right) :=\mu_{D_X}, \ \operatorname{Cov} (D(X_i)) := \Sigma_{D_X},\\
  \mathbb{E} (D(Y_j)) = & \left(
    \begin{matrix}
    \Vert \mathbf{A} \Vert_F^2 \sigma^2(F_x) + \Vert \mathbf{B} \Vert_F^2 \sigma^2(F_y) + \Vert \mu_Y - \mu_X \Vert_2^2 \\
    2\Vert \mathbf{B} \Vert_F^2 \sigma^2(F_y)
    \end{matrix} \right) :=\mu_{D_Y},\ \operatorname{Cov} (D(Y_j)) := \Sigma_{D_Y}.
\end{align*}

\end{theorem}

For a test observation $W$, the distance profile $D(W)$ is defined analogously. 
If $W\sim F_X$, then $\mathbb E\{D(W)\}=\mu_{D_X}$; if $W\sim F_Y$, then 
$\mathbb E\{D(W)\}=\mu_{D_Y}$. The corresponding covariance matrices, $\Sigma_{D_{W_x}}$ and $\Sigma_{D_{W_y}}$, are given 
in Appendix~\ref{appendix:EV}.

The expectations show how distance profiles reflect both location and scale 
differences. The between-class component depends on the mean difference 
$\|\mu_Y-\mu_X\|_2^2$, while the within- and between-class components depend on 
the scale terms $\|\mathbf A\|_F^2\sigma^2(F_x)$ and 
$\|\mathbf B\|_F^2\sigma^2(F_y)$. Higher-order features of the marginal 
distributions enter through the covariance matrices $\Sigma_{D_X}$ and 
$\Sigma_{D_Y}$.

To obtain a distributional reference for the distance-based surrogate, we
study the asymptotic distribution of an individual distance profile. We
partition its coordinate contributions into blocks and retain interiors
separated by gaps so that the retained sums are independent under (C1).
Let $\mathbf a_k^\top$ and $\mathbf b_k^\top$ be the $k$th rows of
$\mathbf A$ and $\mathbf B$, let $\Delta=\mu_Y-\mu_X$, and set
$s=\lfloor d^\alpha\rfloor$, $t=\lfloor d/s\rfloor$, and $r=d-st$, where
$0<\alpha<1/2$.  For $q\in[t]$, define
\begin{align*}
R_q^{(X)}
&=\sum_{k=(q-1)s+1}^{qs-2m_0}
\begin{pmatrix}
\displaystyle\frac{1}{n-1}\sum_{\substack{1\le\ell\le n\\\ell\ne1}}
\left[(\mathbf a_k^\top x_\ell-\mathbf a_k^\top x_1)^2
-\mathbb E[(\mathbf a_k^\top x_\ell-\mathbf a_k^\top x_1)^2]\right]\\[2mm]
\displaystyle\frac{1}{m}\sum_{\ell=1}^m
\left[(\mathbf b_k^\top y_\ell-\mathbf a_k^\top x_1+\Delta_k)^2
-\mathbb E[(\mathbf b_k^\top y_\ell-\mathbf a_k^\top x_1+\Delta_k)^2]\right]
\end{pmatrix},\\
R_q^{(Y)}
&=\sum_{k=(q-1)s+1}^{qs-2m_0}
\begin{pmatrix}
\displaystyle\frac{1}{n}\sum_{\ell=1}^n
\left[(\mathbf a_k^\top x_\ell-\mathbf b_k^\top y_1-\Delta_k)^2
-\mathbb E[(\mathbf a_k^\top x_\ell-\mathbf b_k^\top y_1-\Delta_k)^2]\right]\\[2mm]
\displaystyle\frac{1}{m-1}\sum_{\substack{1\le\ell\le m\\\ell\ne1}}
\left[(\mathbf b_k^\top y_\ell-\mathbf b_k^\top y_1)^2
-\mathbb E[(\mathbf b_k^\top y_\ell-\mathbf b_k^\top y_1)^2]\right]
\end{pmatrix}.
\end{align*}
Let
\[
\Sigma_t^{(X)}=\operatorname{Cov}\!\left(\sum_{q=1}^tR_q^{(X)}\right),
\qquad
\Sigma_t^{(Y)}=\operatorname{Cov}\!\left(\sum_{q=1}^tR_q^{(Y)}\right).
\]
Let $W_x=\mathbf A w_x+\mu_X$ and $W_y=\mathbf B w_y+\mu_Y$,
where $w_x$ and $w_y$ are independent of the training samples and have
i.i.d.\ coordinates from $F_x$ and $F_y$, respectively.
For these test observations,
define $R_{q,W}^{(X)}$ and $R_{q,W}^{(Y)}$ from $R_q^{(X)}$ and $R_q^{(Y)}$
by replacing the anchor $x_1$ or $y_1$ with its test counterpart $w_x$ or $w_y$
and averaging over all training observations in each class.
Let $\Sigma_{t,W}^{(g)}=\operatorname{Cov}(\sum_{q=1}^t R_{q,W}^{(g)})$
for $g\in\{X,Y\}$.

The conditions used in Theorem~\ref{thm:distance_clt} are:
\begin{itemize}[itemsep=1pt,topsep=2pt,parsep=0pt]
\item[(C1)] $\mathbf A$ and $\mathbf B$ are banded with fixed bandwidth
$m_0\in\mathbb N$: $a_{ij}=b_{ij}=0$ whenever $|i-j|\ge m_0$.
\item[(C2)] For fixed constants $\mu_0,c_0$,
$\max_k|\Delta_k|<\mu_0$ and
$\max_{i,j}\{|a_{ij}|,|b_{ij}|\}<c_0$.
\item[(C3)] The independent block sums satisfy
\[
\sum_{q=1}^t\mathbb E\| (\Sigma_t^{(X)})^{-1/2}R_q^{(X)}\|_2^3\to0,
\qquad
\sum_{q=1}^t\mathbb E\| (\Sigma_t^{(Y)})^{-1/2}R_q^{(Y)}\|_2^3\to0.
\]
\item[(C4)] For some $c>0$ and all sufficiently large $d$,
$\lambda_{\min}(\Sigma_t^{(X)})\ge cd$ and
$\lambda_{\min}(\Sigma_t^{(Y)})\ge cd$.
\end{itemize}
For the test-profile conclusions, Conditions~(C3)--(C4) are imposed with
$R_{q,W}^{(g)}$ and $\Sigma_{t,W}^{(g)}$ in place of their training-profile
counterparts.
The proof is given in Appendix~\ref{appB}.
\begin{theorem}
\label{thm:distance_clt}
As $d\to\infty$, with $n,m\geq2$ fixed, under
Setting~$(\star)$ and Conditions~(C1)--(C4),
\[
\left(\Sigma_t^{(X)}\right)^{-1/2}\{D(X_i)-\mu_{D_X}\}
\ \xrightarrow{d}\ N_2(0,I_2),
\qquad
\left(\Sigma_t^{(Y)}\right)^{-1/2}\{D(Y_j)-\mu_{D_Y}\}
\ \xrightarrow{d}\ N_2(0,I_2).
\]
If, in addition, \(d^{-1}\Sigma_t^{(X)}\to V_X\) and
\(d^{-1}\Sigma_t^{(Y)}\to V_Y\) for positive-definite matrices \(V_X,V_Y\), then
\[
\frac{D(X_i)-\mu_{D_X}}{\sqrt d}\xrightarrow{d}N_2(0,V_X),
\qquad
\frac{D(Y_j)-\mu_{D_Y}}{\sqrt d}\xrightarrow{d}N_2(0,V_Y).
\]
If the analogues of Conditions~(C3)--(C4) hold for the test-profile blocks,
then
\[
(\Sigma_{t,W}^{(X)})^{-1/2}\{D(W_x)-\mu_{D_X}\}\xrightarrow{d}N_2(0,I_2),
\qquad
(\Sigma_{t,W}^{(Y)})^{-1/2}\{D(W_y)-\mu_{D_Y}\}\xrightarrow{d}N_2(0,I_2).
\]
The corresponding $d^{-1/2}$-scaled limits follow when
$d^{-1}\Sigma_{t,W}^{(g)}$ converges to a positive-definite matrix.
\end{theorem}

\begin{figure}[!htbp]
\centering
\newcommand{\BayesPanel}[1]{\resizebox{0.47\textwidth}{!}{\setlength{\unitlength}{1bp}\begin{picture}(388,309)
      \put(0,0){\includegraphics[width=388bp]{#1}}
      \put(249,248){\pdfliteral direct{q 1 g}\rule{126bp}{25bp}\pdfliteral direct{Q}}
      \put(250.328,264){\sffamily\fontsize{11}{12}\selectfont Gaussian Bayes error}
      \put(250.328,251.75){\sffamily\fontsize{11}{12}\selectfont QDA test error}
    \end{picture}}}
\BayesPanel{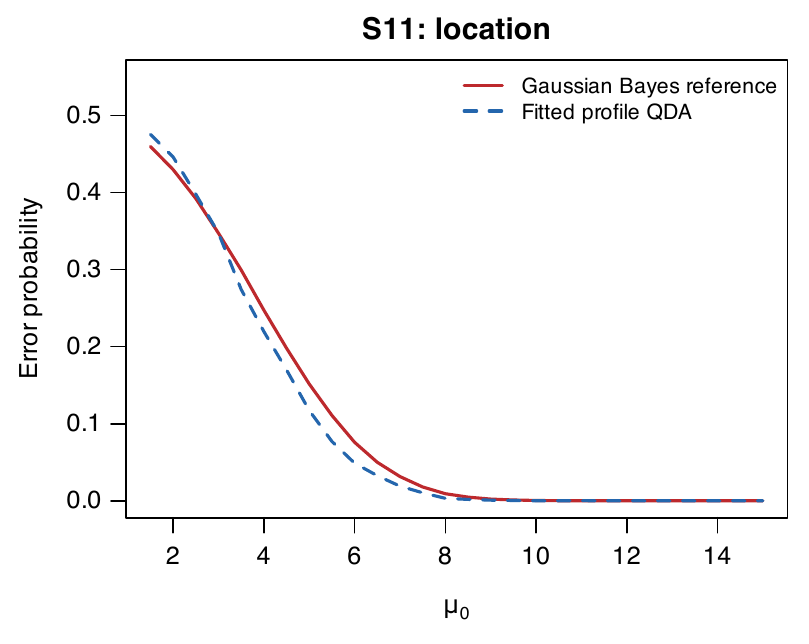}\hfill
\BayesPanel{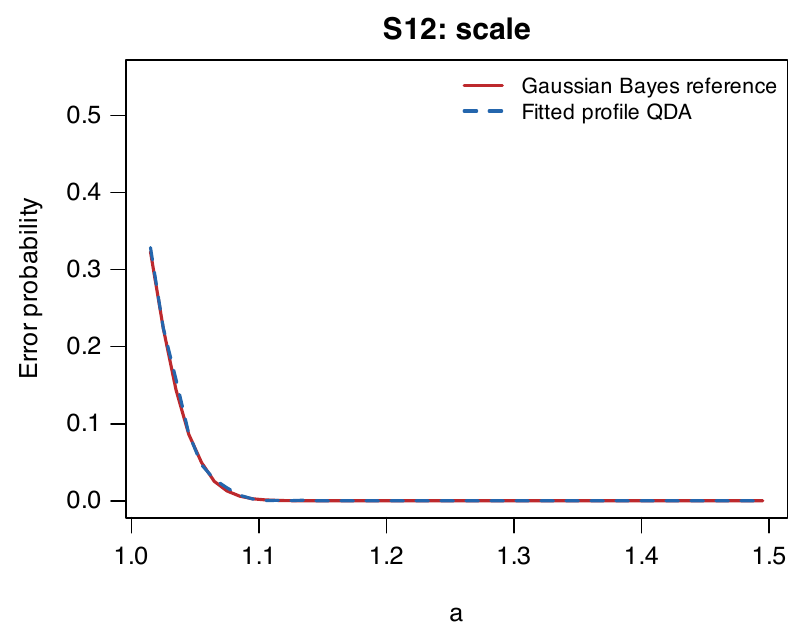}
\caption{Bayes error under the Gaussian approximation (red)
and simulated test error of distance-profile QDA (blue) for Scenario S11
(left) and Scenario S12 (right).}
\label{fig:bayes}
\end{figure}

Theorem~\ref{thm:distance_clt} motivates a Gaussian approximation
to the class-conditional distributions of the test distance profile
$D(W)=(D_X(W),D_Y(W))^\top$:
\[
D(W)\mid G_W=X \mathrel{\dot\sim} N(\mu_{D_X},\Sigma_{D_{W_x}}),
\qquad
D(W)\mid G_W=Y \mathrel{\dot\sim} N(\mu_{D_Y},\Sigma_{D_{W_y}}).
\]
Here $G_W\in\{X,Y\}$ is the class label. We use the Bayes error under this
approximation as a benchmark for the simulated test error of distance-profile QDA.
Under equal priors, the
Bayes error for the approximating profile distributions is
\[
R_{\mathrm G}^*=\int \min\{P(G_W=X\mid D(W)=t),P(G_W=Y\mid D(W)=t)\}
p_{D(W)}(t)\,dt.
\]
Define the Gaussian discriminant scores
\begin{align*}
\delta_X(t)&=-\tfrac12\log|\Sigma_{D_{W_x}}|
-\tfrac12(t-\mu_{D_X})^\top\Sigma_{D_{W_x}}^{-1}(t-\mu_{D_X}),\\
\delta_Y(t)&=-\tfrac12\log|\Sigma_{D_{W_y}}|
-\tfrac12(t-\mu_{D_Y})^\top\Sigma_{D_{W_y}}^{-1}(t-\mu_{D_Y}).
\end{align*}
Then
\[
P(G_W=X\mid D(W)=t)
=\frac{\exp\{\delta_X(t)\}}
{\exp\{\delta_X(t)\}+\exp\{\delta_Y(t)\}},
\]
with the class-$Y$ probability given by its complement. Because $R_{\mathrm G}^*$ has
no closed form in general, we estimate it by averaging the smaller posterior
probability over Monte Carlo draws from
\[
\tfrac12N(\mu_{D_X},\Sigma_{D_{W_x}})
+\tfrac12N(\mu_{D_Y},\Sigma_{D_{W_y}}).
\]
The same test-profile covariance matrices are used both to sample this
mixture and to evaluate its posterior probabilities. Separately, we generate
training and test observations from $F_X,F_Y$, form squared-Euclidean
distance profiles using Algorithm~\ref{dist_algorithm}, fit QDA to the
training profiles, and record the fraction of incorrect test predictions.
Appendix~\ref{app:bayes-computation} gives the simulation settings for Figure~\ref{fig:bayes}.

Figure~\ref{fig:bayes} shows that the Bayes error under the Gaussian
approximation closely tracks the simulated test error of distance-profile QDA.
Both errors fall rapidly as
location separation increases in S11 or scale separation increases in S12.
Both experiments use Gaussian data with AR covariance, $d=2000$, and $n=m=50$.

\section{Discussion}
\label{sec:conclusion}

Class-wise dissimilarity profiles provide a common way to
classify high-dimensional vectors and non-Euclidean data. In S5, the rank-based classifier
had the lowest test error when the classes differed only in correlation
structure, with identical means and marginal variances. In the simulated
graph experiments, it had the lowest error in every setting when all three
classifiers used the same edge-count $L_1$ dissimilarity. Its test error also
increased less than that of the distance-based surrogate under the training
contamination in Table~\ref{outres}.

\textbf{Limitations.}
The choice of dissimilarity matters. When only a few coordinates
carry signal, distances computed over all coordinates can dilute it.
Appendix~\ref{app:supervised_geometry} illustrates how feature screening based
only on the training data can help. For $N$ training vectors of dimension $d$,
computing all pairwise dissimilarities costs $O(N^2d)$. Further work could
develop faster approximations for large samples. Our two-class theory explains
how distributional differences can separate the class-wise profiles of the
distance-based surrogate. The classification performance and robustness to
extreme distances of the rank-based Algorithm~\ref{rank_algorithm} are
assessed empirically.

\textbf{Broader impact.}
High-dimensional classification arises in fields such as genomics,
biomedicine, and network analysis, where labeled data are often limited.
By capturing differences in location, scale, and other distributional features,
our method can help researchers identify class structure without knowing in advance
which differences drive the separation. This flexibility is particularly useful
in exploratory studies. As with other classifiers, use in sensitive applications
calls for careful validation and domain-specific interpretation.

\paragraph{AI use statement}
Generative AI tools assisted with writing and editing,
code development and debugging, experiment design and analysis, and checking mathematical derivations.
The authors reviewed the resulting text, code, and mathematical arguments
and take responsibility for the final content.

\paragraph{Reproducibility statement}
Section~\ref{sec:method} gives the complete algorithm; Section~\ref{sec:simulation}
and the appendix document the experiments and baseline configurations; and the
appendix provides all proofs.

\begin{ack}
The authors were supported in part by NSF awards DMS-1848579 and DMS-2311399.
\end{ack}

\bibliographystyle{plainnat}
\bibliography{myref}    

\appendix
\raggedbottom
\numberwithin{table}{section}
\numberwithin{figure}{section}
\numberwithin{algorithm}{section}

\section{Moments of distance profiles}\label{app_sec4}
\subsection{Training profiles}

\begin{definition}
\label{def1}
Let $r,s,u\in\mathbb R^d$ be mutually independent random vectors whose
coordinates are i.i.d. from the centered distributions $F_r,F_s,F_u$,
respectively.  For matrices $\mathbf C,\mathbf D,\mathbf G\in\mathbb R^{d\times d}$
and vectors $\mu_1,\mu_2,\mu_3\in\mathbb R^d$, define
\begin{align*}
&h_1(\mathbf C,\mathbf D,\mathbf G,\mu_1,\mu_2,\mu_3,F_r,F_s,F_u)
=\mathbb E\!\left[
\|\mathbf D s+\mu_2-(\mathbf C r+\mu_1)\|_2^2
\|\mathbf G u+\mu_3-(\mathbf C r+\mu_1)\|_2^2
\right],\\
&h_2(\mathbf C,\mathbf D,\mu_1,\mu_2,F_r,F_s)
=\mathbb E\!\left[
\|\mathbf D s+\mu_2-(\mathbf C r+\mu_1)\|_2^4
\right].
\end{align*}
These expectations are finite under Setting~$(\star)$.  Because their
integrands are polynomials of total degree at most four in independent
centered coordinates, they depend only on the first four centered moments of
$F_r,F_s,F_u$ and on the displayed matrices and mean vectors.
\end{definition}

\begin{align*}
&h_1(\mathbf{C}, \mathbf{D}, \mathbf{G}, \mu_1, \mu_2, \mu_3, F_r, F_s, F_u)
=\mathbb E\!\left[
\|\mathbf D s+\mu_2-(\mathbf C r+\mu_1)\|_2^2
\|\mathbf G u+\mu_3-(\mathbf C r+\mu_1)\|_2^2
\right]\\
&= \Vert \mathbf{D} \Vert_F^2 \sigma^2(F_s)
   \Vert \mathbf{G} \Vert_F^2 \sigma^2(F_u) + \Vert \mathbf{G} \Vert_F^2 \sigma^2(F_u)
  \Vert \mu_2 - \mu_1 \Vert_2^2
+ \Vert \mathbf{D} \Vert_F^2 \sigma^2(F_s)
  \Vert \mu_3 - \mu_1 \Vert_2^2\notag\\
&\quad
+ \sum_{i=1}^d \sum_{j=1}^d
\left[
\sum_{k=1}^d c_{ik}^2 c_{jk}^2 \kappa(F_r)
+
\left(
\sum_{k_1 \neq k_2} c_{ik_1}^2 c_{jk_2}^2
+
2\sum_{k_1 \neq k_2}
c_{ik_1}c_{ik_2} c_{jk_1} c_{jk_2}
\right)
(\sigma^2(F_r))^2
\right] \notag\\
&\quad
+ \left(
\Vert \mathbf{D} \Vert_F^2 \sigma^2(F_s)
+ \Vert \mathbf{G} \Vert_F^2 \sigma^2(F_u)
\right)
\Vert \mathbf{C} \Vert_F^2 \sigma^2(F_r) +4\sigma^2(F_r)
(\mu_2-\mu_1)^\top
\mathbf C\mathbf C^\top
(\mu_3-\mu_1) \notag\\
&\quad
-2\mathbb{E}_r\left[
\left\{
(\mu_2-\mu_1)^\top \mathbf C r
+
(\mu_3-\mu_1)^\top \mathbf C r
\right\}
\Vert \mathbf C r \Vert_2^2
\right] \notag\\
&\quad
+ \left(
\Vert \mu_2 - \mu_1 \Vert_2^2
+
\Vert \mu_3 - \mu_1 \Vert_2^2
\right)
\Vert \mathbf{C} \Vert_F^2 \sigma^2(F_r) + \Vert \mu_2 - \mu_1 \Vert_2^2
  \Vert \mu_3 - \mu_1 \Vert_2^2.
\end{align*}

\begingroup
\allowdisplaybreaks[0]
\begin{align*}
    &h_2(\mathbf{C}, \mathbf{D}, \mu_1, \mu_2, F_r, F_s)
    =\mathbb E\!\left[
\|\mathbf D s+\mu_2-(\mathbf C r+\mu_1)\|_2^4 \right]\\
    &= 2 \Vert \mu_2 - \mu_1 \Vert_2^2 \left( \Vert \mathbf{D} \Vert_F^2 \sigma^2(F_s) + \Vert \mathbf{C} \Vert_F^2 \sigma^2(F_r) \right) \\
    &\quad + \sum_{i=1}^d \sum_{j=1}^d \left( \sum_{k=1}^d d_{ik}^2 d_{jk}^2 \kappa(F_s) + \left(\sum_{k_1 \neq k_2} d_{ik_1}^2 d_{jk_2}^2 + 2\sum_{k_1 \neq k_2} d_{ik_1}d_{ik_2} d_{jk_1} d_{jk_2} \right) (\sigma^2(F_s))^2 \right) \\
    &\quad + \sum_{i=1}^d \sum_{j=1}^d \left( \sum_{k=1}^d c_{ik}^2 c_{jk}^2 \kappa(F_r) + \left(\sum_{k_1 \neq k_2} c_{ik_1}^2 c_{jk_2}^2 + 2\sum_{k_1 \neq k_2} c_{ik_1}c_{ik_2} c_{jk_1} c_{jk_2} \right) (\sigma^2(F_r))^2 \right) \\
    &\quad + 4 \sum_{i=1}^d \sum_{j=1}^d \left( \sum_{k=1}^d d_{ik}^2 d_{jk} (\mu_2 - \mu_1)_j \right) \gamma(F_s)  - 4 \sum_{i=1}^d \sum_{j=1}^d \left( \sum_{k=1}^d c_{ik}^2 c_{jk} (\mu_2 - \mu_1)_j \right) \gamma(F_r) \\
    &\quad + 4 \Vert \mathbf{C}^\top(\mu_2 - \mu_1)  \Vert_2^2 \sigma^2(F_r) + 2\Vert \mathbf{D} \Vert_F^2 \sigma^2(F_s) \Vert \mathbf{C} \Vert_F^2 \sigma^2(F_r) + \Vert (\mu_2 - \mu_1) \Vert_2^4 \\
    &\quad + 4 \Vert \mathbf{D}^\top (\mu_2 - \mu_1)  \Vert_2^2 \sigma^2(F_s) + 4 \Vert \mathbf{C}^\top \mathbf{D} \Vert_F^2 \sigma^2(F_s) \sigma^2(F_r).
\end{align*}
Here $\sigma^2(F)=\mathbb E Z^2$, $\gamma(F)=\mathbb E Z^3$, and
$\kappa(F)=\mathbb E Z^4$ for a centered variable $Z\sim F$.
In particular, $\gamma$ and $\kappa$ are unstandardized centered moments,
not standardized skewness and kurtosis. To see the shift-dependent terms,
write $\delta=\mu_2-\mu_1$ and $L=\mathbf D s-\mathbf C r$. Expanding
$\|L+\delta\|_2^4$ gives $4\delta^\top\operatorname{Cov}(L)\delta$
and $4\delta^\top\mathbb E[L\|L\|_2^2]$; independence of the centered
coordinates yields the transposed-matrix and third-moment terms above.

\endgroup

\begin{definition}

\label{def2}

    The covariance matrices $\Sigma_{D_X}$ and $\Sigma_{D_Y}$
    in Theorem~\ref{theorem1} are defined as follows:

\begin{align*}
    \Sigma_{D_X}[1,1] = & \frac{n-2}{n-1} h_1(\mathbf{A}, \mathbf{A}, \mathbf{A}, \mu_X, \mu_X, \mu_X, F_x, F_x, F_x) + \frac{1}{n-1} h_2(\mathbf{A}, \mathbf{A}, \mu_X, \mu_X, F_x, F_x ) \\
    & - 4 \Vert \mathbf{A} \Vert_F^4 (\sigma^2(F_x))^2,\\
     \Sigma_{D_X}[1,2] = & \Sigma_{D_X}[2,1] = h_1(\mathbf{A}, \mathbf{A}, \mathbf{B}, \mu_X, \mu_X, \mu_Y, F_x, F_x, F_y) \\
    & - 2 \Vert \mathbf{A} \Vert_F^2 \sigma^2(F_x)( \Vert \mathbf{B}  \Vert_F^2 \sigma^2(F_y) + \Vert \mathbf{A} \Vert_F^2 \sigma^2(F_x) + \Vert \mu_Y - \mu_X \Vert_2^2 ), \\
    \Sigma_{D_X}[2,2] = & \frac{m-1}{m} h_1(\mathbf{A}, \mathbf{B}, \mathbf{B}, \mu_X, \mu_Y, \mu_Y, F_x, F_y, F_y) + \frac{1}{m} h_2(\mathbf{A}, \mathbf{B}, \mu_X,  \mu_Y, F_x,  F_y) \\ 
    & - ( \Vert \mathbf{B} \Vert_F^2 \sigma^2(F_y) + \Vert \mathbf{A} \Vert_F^2 \sigma^2(F_x) + \Vert \mu_Y - \mu_X \Vert_2^2 )^2, \\
    \Sigma_{D_Y}[1,1] = & \frac{n-1}{n} h_1(\mathbf{B}, \mathbf{A}, \mathbf{A}, \mu_Y, \mu_X, \mu_X, F_y, F_x, F_x) + \frac{1}{n} h_2(\mathbf{B}, \mathbf{A}, \mu_Y,  \mu_X, F_y,  F_x) \\ 
    & - ( \Vert \mathbf{A} \Vert_F^2 \sigma^2(F_x) + \Vert \mathbf{B} \Vert_F^2 \sigma^2(F_y) + \Vert \mu_Y - \mu_X \Vert_2^2 )^2, \\
    \Sigma_{D_Y}[1,2] = & \Sigma_{D_Y}[2,1] = h_1(\mathbf{B}, \mathbf{A}, \mathbf{B}, \mu_Y, \mu_X, \mu_Y, F_y, F_x, F_y) \\
    & - 2 \Vert \mathbf{B} \Vert_F^2 \sigma^2(F_y)( \Vert \mathbf{A} \Vert_F^2 \sigma^2(F_x) + \Vert \mathbf{B} \Vert_F^2 \sigma^2(F_y) + \Vert \mu_Y - \mu_X \Vert_2^2 ), \\
    \Sigma_{D_Y}[2,2] = & \frac{m-2}{m-1} h_1(\mathbf{B}, \mathbf{B}, \mathbf{B}, \mu_Y, \mu_Y, \mu_Y, F_y, F_y, F_y) + \frac{1}{m-1} h_2(\mathbf{B}, \mathbf{B}, \mu_Y, \mu_Y, F_y, F_y ) \\ 
    &- 4 \Vert \mathbf{B} \Vert_F^4 (\sigma^2(F_y))^2.
\end{align*}

\end{definition}

\par\edef\testprofileparskip{\the\parskip}\noindent\begin{minipage}{\textwidth}
\setlength{\parskip}{\testprofileparskip}
\allowdisplaybreaks[0]
\subsection{Test profiles}\label{appendix:EV}

The means and covariance matrices of $D(W_x)$ and $D(W_y)$ are stated in Theorem \ref{theorem4}, with the proof provided in Appendix~\ref{appA}.
\begin{theorem}

\label{theorem4}

    Let the test observations be independent of the training samples.
    Under Setting ($\star$), suppose $W_x \sim F_X$ and $W_y\sim F_Y$. Then we have $ \mathbb{E}\{D(W_x)\}=\mu_{D_X}$ and $\mathbb{E}\{D(W_y)\} =\mu_{D_Y}$, where $\mu_{D_X}$ and $\mu_{D_Y}$ are given in Theorem \ref{theorem1}. The covariance matrices are $\operatorname{Cov}\{D(W_x)\} = \Sigma_{D_{W_x}}$ and $\operatorname{Cov}\{D(W_y)\} = \Sigma_{D_{W_y}}$, where

\begin{align*}
     \Sigma_{D_{W_x}}[1,1] = & \frac{n-1}{n} h_1(\mathbf{A}, \mathbf{A}, \mathbf{A}, \mu_X, \mu_X, \mu_X, F_x, F_x, F_x) + \frac{1}{n} h_2(\mathbf{A}, \mathbf{A}, \mu_X, \mu_X, F_x, F_x ) \\ 
     & - 4 \Vert \mathbf{A} \Vert_F^4 (\sigma^2(F_x))^2,\\
     \Sigma_{D_{W_x}}[1,2] = & \Sigma_{D_{W_x}}[2,1] = h_1(\mathbf{A}, \mathbf{A}, \mathbf{B}, \mu_X, \mu_X, \mu_Y, F_x, F_x, F_y) \\
    & - 2 \Vert \mathbf{A} \Vert_F^2 \sigma^2(F_x)( \Vert \mathbf{B}  \Vert_F^2 \sigma^2(F_y) + \Vert \mathbf{A} \Vert_F^2 \sigma^2(F_x) + \Vert \mu_Y - \mu_X \Vert_2^2 ), \\
     \Sigma_{D_{W_x}}[2,2] = & \frac{m-1}{m} h_1(\mathbf{A}, \mathbf{B}, \mathbf{B}, \mu_X, \mu_Y, \mu_Y, F_x, F_y, F_y) + \frac{1}{m} h_2(\mathbf{A}, \mathbf{B}, \mu_X,  \mu_Y, F_x,  F_y) \\ 
    & - ( \Vert \mathbf{B} \Vert_F^2 \sigma^2(F_y) + \Vert \mathbf{A} \Vert_F^2 \sigma^2(F_x) + \Vert \mu_Y - \mu_X \Vert_2^2 )^2, \\
     \Sigma_{D_{W_y}}[1,1] = & \frac{n-1}{n} h_1(\mathbf{B}, \mathbf{A}, \mathbf{A}, \mu_Y, \mu_X, \mu_X, F_y, F_x, F_x) + \frac{1}{n} h_2(\mathbf{B}, \mathbf{A}, \mu_Y,  \mu_X, F_y,  F_x) \\ 
    & - ( \Vert \mathbf{A} \Vert_F^2 \sigma^2(F_x) + \Vert \mathbf{B} \Vert_F^2 \sigma^2(F_y) + \Vert \mu_Y - \mu_X \Vert_2^2 )^2, \\
     \Sigma_{D_{W_y}}[1,2] = & \Sigma_{D_{W_y}}[2,1] = h_1(\mathbf{B}, \mathbf{A}, \mathbf{B}, \mu_Y, \mu_X, \mu_Y, F_y, F_x, F_y) \\
    & - 2 \Vert \mathbf{B} \Vert_F^2 \sigma^2(F_y)( \Vert \mathbf{A} \Vert_F^2 \sigma^2(F_x) + \Vert \mathbf{B} \Vert_F^2 \sigma^2(F_y) + \Vert \mu_Y - \mu_X \Vert_2^2 ), \\
     \Sigma_{D_{W_y}}[2,2] = & \frac{m-1}{m} h_1(\mathbf{B}, \mathbf{B}, \mathbf{B}, \mu_Y, \mu_Y, \mu_Y, F_y, F_y, F_y) + \frac{1}{m} h_2(\mathbf{B}, \mathbf{B}, \mu_Y, \mu_Y, F_y, F_y ) \\ 
     &- 4 \Vert \mathbf{B} \Vert_F^4 (\sigma^2(F_y))^2.
\end{align*}
\end{theorem}
\end{minipage}\par

\section{Simulation settings for the Gaussian benchmark}
\label{app:bayes-computation}

For the comparison in Figure~\ref{fig:bayes}, we set $n=m=50$ and $d=2000$,
take $\mathbf B=a\mathbf A$ with $\mathbf A\mathbf A^\top=\Sigma$ and
$\Sigma_{r,c}=0.1^{|r-c|}$, and set $\mu_X=0$ and
$\mu_Y=\mu_0\mu'/\|\mu'\|$, where $\mu'\sim N_d(0,\mathbf I_d)$.  Scenario
S11 has $F_x=F_y=N(0,1)$ and $a=1$, with $\mu_0$ varying; Scenario S12 has
$F_x=F_y=N(0,1)$ and $\mu_0=0$, with $a$ varying.
We use $\mu_0=1.5,2.0,\ldots,15.0$ in S11 and
$a=1.015,1.025,\ldots,1.495$ in S12. The mean direction is drawn once and
held fixed. Each setting uses 50 independent training/test repetitions,
with 50 test observations per class in each repetition. For the Gaussian
reference, each repetition uses 500 independent draws from
each of the two class-conditional Gaussian distributions. The curves show averages over repetitions;
Monte Carlo standard errors are retained in the accompanying results.
The dense AR covariance $\Sigma$ does not satisfy (C1): fixed-bandwidth
$\mathbf A$ would imply a banded $\mathbf A\mathbf A^\top$. These
experiments therefore examine the Gaussian reference empirically beyond
the theorem's stated assumptions.

The Gaussian reference averages over random training samples when specifying
the distribution of $D(W)$, whereas fitted QDA uses the realized training
sample. The fitted-error curve can therefore lie below the reference.

\par\noindent\begin{minipage}{\textwidth}
\section{Detailed distance-based algorithm} \label{app_dist}

\begin{algorithm}[H]
\caption{Dissimilarity-Profiling Classification Algorithm}
\label{dist_algorithm}

\begin{itemize}
    \item [1.] Construct the pairwise dissimilarity matrix $D \in \mathbb{R}^{N \times N}$ where $D[i,j] = \Vert Z_i - Z_j \Vert_2^2.$
    \item [2.] Compute the summary statistics matrix $M^{(D)} \in \mathbb{R}^{N \times k}$: $$M^{(D)}[i,j]=  \frac{\sum_{g_l = j, l \neq i} D[i,l]}{n_j - \mathbbm{1} \left \{g_i = j \right \}  }.$$
    \item [3.] For a new observation $W$, compute $D_W = (D_{W,1},D_{W,2},\ldots,D_{W,N}) \in \mathbb{R}^N$, where $ D_{W,i} = \Vert W- Z_i \Vert_2^2. $  Then, compute $M^{(D)}_W =  (M^{(D)}_1(W),M^{(D)}_2(W),\ldots,M^{(D)}_k(W)) \in \mathbb{R}^k$, where $M^{(D)}_i(W) = \sum _{g_l = i} D_{W,l}/n_i.$ 
    \item [4.] Use QDA to classify $M^{(D)}_W$: \small{$$ g^{(D)}(W) = \arg\max_j\biggl\{-\frac{1}{2} \log|\hat{\Sigma}^{(D)}_j| - \frac{1}{2}(M^{(D)}_W-\hat{\mu}_j^{(D)})^{\top}\!(\hat{\Sigma}^{(D)}_j)^{-1}\!(M^{(D)}_W-\hat{\mu}^{(D)}_j) + \log(n_j/N)\biggr\},  $$}
    where $v_i=M^{(D)}[i,\cdot]^\top$ is the $i$th training profile written as a column vector, \\
    $  \hat{\mu}^{(D)}_j = \frac{1}{n_j} \sum_{i=1}^N v_i \mathbbm{1} \left \{g_i =j \right \},$ and \\
    $\hat{\Sigma}^{(D)}_j = \frac{1}{n_j-1} \sum_{i=1}^N \left( v_i- \hat{\mu}^{(D)}_j \right)\left( v_i- \hat{\mu}^{(D)}_j \right)^\top \mathbbm{1} \left \{g_i =j \right \}.$
\end{itemize}

\end{algorithm}
\end{minipage}\par

Tables \ref{tab:dr1} and \ref{tab:dr2} show that in the absence of outliers, both algorithms perform comparably well across a variety of settings. 

\begin{table}[H]
\centering
\footnotesize
\setlength{\tabcolsep}{6pt}
\caption{Two-class misclassification rate for 1,000-dimensional data.}
\label{tab:dr1}
\begin{adjustbox}{max width=\textwidth}
\begin{tabular}{lcccccccccc}
\hline
Scenario & S1 & S1 & S1 & S2 & S2 & S2 & S3 & S3 & S3 & S4 \\
\hline
$\mu_0$ & 6 & 0 & 6 & 6 & 0 & 6 & 6 & 0 & 6 & 0 \\
\hline
$a$ & 1 & 1.1 & 1.1 & 1 & 1.1 & 1.1 & 1 & 1.1 & 1.1 & 1 \\
\hline
Alg.~\ref{rank_algorithm} & 0.023 & 0.019 & 0.004 & 0.109 & 0.118 & 0.044 & 0.428 & 0.084 & 0.084 & 0.000 \\
Alg.~\ref{dist_algorithm} & 0.022 & 0.019 & 0.003 & 0.095 & 0.120 & 0.043 & 0.426 & 0.083 & 0.083 & 0.000 \\
\hline
\end{tabular}
\end{adjustbox}
\end{table}

\begin{table}[H]
\centering
\footnotesize
\setlength{\tabcolsep}{6pt}
\caption{Multi-class misclassification rates for 1,000-dimensional data.}
\label{tab:dr2}
\begin{adjustbox}{max width=\textwidth}
\begin{tabular}{lccc}
\hline
Scenario & S6 & S7 & S8 \\
\hline
Alg.~\ref{rank_algorithm} & 0.023 & 0.139 & 0.227 \\
Alg.~\ref{dist_algorithm} & 0.023 & 0.138 & 0.216 \\
\hline
\end{tabular}
\end{adjustbox}
\end{table}

\section{Robustness to training-sample contamination}
\label{appendix_outlier}

\begingroup
We compare the rank- and distance-profile algorithms in four focused
training-contamination settings, holding class-wise averaging and QDA
fixed. Both use leave-one-out training-class averages; Distance uses
squared Euclidean dissimilarities. Both methods use the same generated
data over 50 repetitions per setting. Only class-$X$ training observations
are contaminated; test observations remain clean.

\paragraph{OA--OC: extreme-distance contamination.}
These settings share a $d=200$ Gaussian location model,
$X\sim N_d(0,I_d)$ and $Y\sim N_d(\mu,I_d)$, with
$\mu=(3/\sqrt d)\mathbf 1_d$. Each class has 50 training and
100 independent clean test observations. We replace
$n_o\in\{1,3,5\}$ class-$X$ training observations using:
\begin{itemize}
    \item OA (shift and scale): $X^O_\ell=5Z_\ell+5\mu$;
    \item OB (heavy-tailed rows): $X^O_\ell=2C_\ell$;
    \item OC (single-coordinate spikes): $X^O_\ell=Z_\ell+50S_\ell e_\ell$.
\end{itemize}
Here $Z_\ell\sim N_d(0,I_d)$, $C_\ell$ has independent standard-Cauchy
coordinates, $S_\ell$ is uniform on $\{-1,1\}$, and $e_\ell$ is the
$\ell$th coordinate vector. The shared clean baseline for OA--OC differs
from OD in dimension, mean shift, covariance, and test size.

\paragraph{OD: amplified location contamination.}
Clean observations are $X_i=\mathbf A x_i$ and $Y_j=\mathbf A y_j+\mu$,
where $x_i,y_j\stackrel{\mathrm{i.i.d.}}{\sim}N_d(0,\mathbf I_d)$,
$\mathbf A\mathbf A^\top=\Sigma$, $\Sigma_{r,c}=0.1^{|r-c|}$, and
$d=1000$. We set $\mu=6\mu'/\|\mu'\|_2$, with
$\mu'\sim N_d(0,\mathbf I_d)$ redrawn each repetition and shared by
both methods. Each class has 50 training and 50 independent test
observations. We replace $n_o\in\{0,3,5,7\}$ class-$X$ training
observations by $X^O_\ell=\mathbf A x^O_\ell+5\mu$, where
$x^O_\ell\stackrel{\mathrm{i.i.d.}}{\sim}N_d(0,\mathbf I_d)$.

\begin{table}[H]
\centering
\footnotesize
\setlength{\tabcolsep}{6pt}
\caption{Mean test error (Monte Carlo SE) under training-sample contamination,
over 50 repetitions. OA--OC and OD use the distinct clean models specified
above, with separate clean baselines.}
\label{outres}
\begin{adjustbox}{max width=\textwidth}
\begin{tabular}{llccc}
\hline
Scenario & Contamination & $n_o$ & Rank profile & Distance profile \\
\hline
OA--OC & None & 0 & 0.149 (0.003) & 0.145 (0.003) \\
OA & Shift and scale & 1 & 0.142 (0.004) & 0.361 (0.010) \\
 & & 3 & 0.169 (0.005) & 0.463 (0.005) \\
 & & 5 & 0.179 (0.005) & 0.484 (0.002) \\
OB & Cauchy rows & 1 & 0.148 (0.004) & 0.495 (0.004) \\
 & & 3 & 0.159 (0.004) & 0.500 ($<0.001$) \\
 & & 5 & 0.169 (0.005) & 0.500 (0.000) \\
OC & Coordinate spike & 1 & 0.148 (0.004) & 0.271 (0.008) \\
 & & 3 & 0.150 (0.004) & 0.349 (0.009) \\
 & & 5 & 0.164 (0.004) & 0.404 (0.007) \\
\hline
OD & None & 0 & 0.024 (0.002) & 0.023 (0.002) \\
 & Amplified location & 3 & 0.036 (0.002) & 0.076 (0.004) \\
 & & 5 & 0.055 (0.004) & 0.155 (0.007) \\
 & & 7 & 0.069 (0.004) & 0.270 (0.009) \\
\hline
\end{tabular}
\end{adjustbox}
\end{table}

Table~\ref{outres} shows similar clean-data performance within each model.
Under OA--OC, rank-profile error changes comparatively modestly while
distance-profile error increases sharply. In OD, distance-profile error
increases from 0.023 to 0.270, whereas rank-profile error increases from
0.024 to 0.069. All fits completed successfully.
\endgroup

\section{Additional simulation studies}
\label{app:additional_simulations}

\begingroup
\subsection{Covariance-structure experiments}
\label{app:covariance_pilots}

\subsubsection{Correlation-strength pilot}
To assess how performance changes with the strength of the dependence
difference, we also ran a focused correlation pilot over
$\rho\in\{0.5,0.6,0.7,0.8,0.9\}$, replacing $\Sigma_{0.9}$ in S5 by
$\Sigma_\rho$. This pilot used $d=1000$, 200 training observations per class,
50 test observations per class, and 50 repetitions. Table~\ref{tab:correlation_rho_pilot} reports the resulting mean
misclassification rates.

\begin{table}[H]
\centering
\footnotesize
\setlength{\tabcolsep}{6pt}
\caption{Mean misclassification rate in the focused correlation-strength pilot.}
\label{tab:correlation_rho_pilot}
\begin{adjustbox}{max width=\textwidth}
\begin{tabular}{lccccc}
\hline
$\rho$ & 0.5 & 0.6 & 0.7 & 0.8 & 0.9 \\
\hline
Rank profile & 0.386 & 0.325 & 0.260 & 0.182 & 0.122 \\
Distance profile & 0.433 & 0.386 & 0.350 & 0.281 & 0.189 \\
\hline
\end{tabular}
\end{adjustbox}
\end{table}

The error decreases for both procedures as the dependence difference grows,
while the rank profile is uniformly more accurate across the five values of
$\rho$.

\subsubsection{Equal-spectrum signed-block covariance pilot}
We also consider a more stringent covariance-only experiment with
$X\sim N_d(0,\Sigma_X)$ and $Y\sim N_d(0,\Sigma_Y)$, where $d=1000$ is
partitioned into 20 independent blocks of size 50. Within each block,
\[
\Sigma_X^{(b)}=(1-\rho)I+\rho uu^\top,
\qquad
\Sigma_Y^{(b)}=(1-\rho)I+\rho ss^\top,
\]
where $u$ is the all-one vector and $s$ has 25 entries equal to $+1$ and 25
entries equal to $-1$. For $\rho>0$, $X$ has uniformly positive within-block
correlations, whereas $Y$ has positive correlations within two subgroups and
negative correlations between them. The two covariance matrices have identical
marginal variances, traces, determinants, Frobenius norms, and eigenvalues; they
differ only in their eigenspaces and signed dependence patterns. Using 50
training and 50 test observations per class over 50 repetitions gives the results in Table~\ref{tab:equal_spectrum_signed_block}.

\begin{table}[H]
\centering
\footnotesize
\setlength{\tabcolsep}{2.5pt}
\caption{Mean misclassification rate (Monte Carlo SE) in the equal-spectrum signed-block covariance experiment.}
\label{tab:equal_spectrum_signed_block}
\begin{adjustbox}{max width=\textwidth}
\begin{tabular}{lccccccc}
\hline
$\rho$ & 0.00 & 0.10 & 0.20 & 0.30 & 0.50 & 0.70 & 0.90 \\
\hline
Rank profile & 0.504 (0.007) & 0.458 (0.006) & 0.314 (0.008) & 0.252 (0.007) & 0.186 (0.006) & 0.161 (0.006) & 0.150 (0.007) \\
Distance profile & 0.500 (0.007) & 0.499 (0.006) & 0.492 (0.007) & 0.486 (0.007) & 0.485 (0.007) & 0.469 (0.008) & 0.478 (0.008) \\
\hline
\end{tabular}
\end{adjustbox}
\end{table}

At $\rho=0$ the distributions are identical and both procedures perform at
chance. As the signed dependence difference increases, rank-profile error
decreases steadily, while the raw-distance profile remains close to chance.
\endgroup

\subsection{Training-only supervised-geometry diagnostics}
\label{app:supervised_geometry}

We examine whether training-only feature selection can help when only a few coordinates are informative. In two
$d=100$ settings, four coordinates carry signal and the remaining 96 are
independent $N(0,1)$ noise.  The sparse-mean classes differ by $-0.5$ versus
$+0.5$ on each signal coordinate.  In the equal-trace variance swap, both
classes have mean zero and their signal-coordinate standard deviations are
$(1.5,1.5,0.5,0.5)$ and $(0.5,0.5,1.5,1.5)$.  Using training labels only, we
select four coordinates by either an absolute Welch score or a fixed score
combining standardized mean and log-variance differences, then compute the
usual rank profiles in the selected space.  The oracle column supplies the
four true coordinates but still uses the rank-profile classifier.  Table
\ref{tab:supervised_geometry} reports 50-repetition mean test error and Monte
Carlo SE.

\begingroup
For each regime in Table~\ref{tab:supervised_geometry}, we used 50 training and
100 independent test observations per class in each of 50 repetitions. The absolute
Welch score for coordinate $j$ was
\[
S_j^{(\mathrm W)}=
\frac{|\bar x_{1j}-\bar x_{0j}|}
{\sqrt{\widehat\sigma_{1j}^2/n_1+\widehat\sigma_{0j}^2/n_0}},
\]
and the mean/variance score was
\[
S_j^{(\mathrm{MV})}=
\frac{|\bar x_{1j}-\bar x_{0j}|}
{\sqrt{(\widehat\sigma_{1j}^2+\widehat\sigma_{0j}^2)/2}}+
\left|\log\frac{\max(\widehat\sigma_{1j}^2,10^{-8})}
{\max(\widehat\sigma_{0j}^2,10^{-8})}\right|.
\]
For either score, the four coordinates with the highest
scores computed from the training data define the binary diagonal geometry
\[
d_{\widehat w}^2(x,x')=\sum_{j=1}^d\widehat w_j(x_j-x_j')^2,
\qquad \widehat w_j\in\{0,1\}.
\]
No test observation is used to select the coordinates.  Across the full-space,
learned, and oracle columns, the rank-profile construction and QDA rule are the
same; only the coordinates entering the dissimilarity change.  Random forest
uses all 100 original coordinates, 300 trees, and otherwise default settings.

\begin{table}[H]
\centering
\footnotesize
\setlength{\tabcolsep}{4pt}
\caption{Sparse-signal diagnostics: mean test error with Monte Carlo SE in parentheses. Learned geometries use training labels only.}
\label{tab:supervised_geometry}
\begin{adjustbox}{max width=\textwidth}
\begin{tabular}{lccccc}
\hline
Regime & Full-space & Learned (Welch) & Learned (mean/variance) & Oracle 4D & RF \\
\hline
Sparse mean & 0.247 (0.006) & 0.169 (0.004) & 0.204 (0.006) & 0.164 (0.004) & 0.231 (0.006) \\
Equal-trace variance swap & 0.496 (0.005) & 0.482 (0.008) & 0.280 (0.008) & 0.280 (0.008) & 0.240 (0.005) \\
\hline
\end{tabular}
\end{adjustbox}
\end{table}

Both screening rules retain four coordinates, the number
of signal coordinates in these simulations. Welch screening nearly recovers
the oracle result in the sparse-mean setting. In the variance-swap setting,
screening by both mean and variance matches the oracle result of $0.280$,
whereas Welch screening does not. The remaining error shows that removing
the noise coordinates alone does not resolve this alternative for the
rank-profile classifier.

\paragraph{Relation to established supervised dimension reduction.}
Possible extensions include sure independence screening for ultrahigh-dimensional
feature filtering~\citep{fan2008sure}, supervised principal components~\citep{bair2006prediction}, sparse discriminant analysis~\citep{clemmensen2011sparse}, and large-margin nearest-neighbor metric learning~\citep{weinberger2009distance}. These methods can learn a representation or
metric before classification and are natural extensions when signal is
localized.
With few labeled training observations, regularization helps
prevent overfitting when learning a supervised subspace or metric.
\endgroup

\section{Hyperparameter settings}
\label{app:hyperparam}

The experiments use R, with TensorFlow for MLP training.
Unless specified below, we use the software defaults.

\paragraph{Rank and distance profiles.}
The final classifier is quadratic discriminant analysis as implemented by
\texttt{MASS::qda}; no validation or tuning is applied after constructing the
class-wise profile.

\paragraph{Generalized Linear Discriminant Analysis (GLDA).}
For $p>n$, we use the generalized discriminant subspace of
\mbox{\citet{li2005robust}} and classify by the nearest projected class mean.  When
$p\leq n$, we use \texttt{MASS::lda}.

\paragraph{Support Vector Machine (SVM).}
We use the \texttt{e1071} package with the radial basis function kernel and default parameter settings.
For the simulated-network experiments, we instead use
\texttt{kernlab::ksvm} with the precomputed kernel
$K(G,G')=\exp\{-d_{\mathrm{net}}(G,G')^2/(2\hat\sigma^2)\}$, where
$\hat\sigma$ is the median positive training-pair dissimilarity. The same
training-derived bandwidth is used for the test kernel.
On PROTEINS, the radial SVM uses these defaults after
training-only removal of constant features, including the package's default
feature scaling. The separately reported WL SVM instead uses a linear kernel
on the normalized WL2 features, with cost parameter 1.

\paragraph{Random Forest (RF).}
We use the \texttt{randomForest} package with its default 500 trees and feature
subsampling.  The supervised-geometry diagnostic in
Appendix~\ref{app:supervised_geometry} uses 300 trees, as stated there.

\paragraph{glmnet.}
We use \texttt{cv.glmnet} to select the regularization parameter via cross-validation. Logistic regression is used for binary classification.
We call \texttt{cv.glmnet} with \texttt{nfolds=10}, using
binomial or multinomial response as appropriate. Nutt has only eight training
observations per split: its default fold assignment has eight nonempty
validation groups, each holding out one observation, rather than ten
nonempty folds. The associated small-sample warnings are retained in the logs.

\paragraph{LightGBM.}
We use the \texttt{lightgbm} package with learning rate 0.05, number of leaves 31, feature fraction 0.9, bagging fraction 0.9, and minimum data in leaf 20. Early stopping is applied using a validation split (20\% of the training data).
All rerun tables use at most 5,000 rounds and early stopping
after 50 rounds without validation improvement, using binary or multiclass
log loss. Labels are encoded from zero; multiclass predictions use the
largest probability in each observation's class-probability row.
With very small training sets, the fixed minimum leaf size can prevent
tree splitting.

\paragraph{gMADD.}
The gMADD classifier is implemented using the transformation functions
\[
\gamma(t) = 1 - \exp(-t), \quad \phi(t) = t,
\]
with the nearest-training-observation rule used in the vector experiments.
On PROTEINS, the gMADD-style WL2 distance-row comparator uses five
neighbors and the same normalized WL2 Euclidean dissimilarity.
For the simulated graphs, gMADD uses mean absolute
differences between distance rows for 5-nearest-neighbor classification.

\paragraph{RP-Ensemble.}
We use the \texttt{RPEnsemble} package with projection
dimension $d=\min(5,p)$. The ensemble aggregates $B_1=50$ selected
projections, each chosen from a block of $B_2=10$ candidate projections by
training error. Linear discriminant analysis is the base classifier.

\paragraph{Multilayer perceptron (MLP).}
The network has three hidden layers of 1,000 units with GELU activations and a
softmax output layer.  It is optimized with Adam for 10 epochs using batches of
32 observations. The numerical tables use deterministic CPU TensorFlow
2.15.1 through R, with explicit seeds and Adam parameters: learning rate
0.001, $\beta_1=0.9$, $\beta_2=0.999$, and $\epsilon=10^{-7}$.
Appendix~\ref{appD} varies activation or total training size as specified.

\paragraph{Rank distribution classifier (RDC).}
We use the empirical plug-in rule corresponding to
Equation~(2) of \mbox{\citet{makinde2020rank}}, with the strict empirical CDF used in
the authors' supplementary implementation. For each observation, we compute
the spatial-rank outlyingness percentile relative to each class and select
the class with the smallest score. For ties among the minimizing classes,
we report the expected error under uniform random tie-breaking; a binary
tie therefore receives half loss. The same rule is used in the two-class
and multi-class experiments.

\section{Computational resources}
\label{app:computing}

The original experiments used a Linux server with two Intel
Xeon E5-2699 v3 processors (36 physical cores, 72 logical threads) and about
264 GB RAM. The numerical rerun used R 4.4.0 on Rocky Linux 8.10, with 40 CPU
workers and two MLP workers in separate stages and internal threads limited
to one. Python 3.11.17, TensorFlow 2.15.1, NumPy 1.26.4, and a compatible
legacy Keras backend were recorded. Across the initial full rerun and the
subsequent gene-expression and outlier reruns, the cumulative batch
wall-clock time was approximately 2.7 hours. Each batch was timed from its
first saved task's start to its last saved task's completion, excluding
installation, pre-run data preparation, and idle time between batches.

The multi-class RDC comparison used ten CPU workers, with
the same generator and simulation seeds as the other methods.

The simulations use 50 repetitions per setting. A master seed
of 20261021 determines the cell and repetition seeds; paired
methods use the same data and separate method seeds. Gene-expression
candidate seeds start at 1 and increase sequentially until 50 distinct
QDA-admissible splits have been selected by the training-only rule in
Section~\ref{sec:nutt_counts}.
For PROTEINS, we use seeds 60218001--60218050. The
setup for the Gaussian comparison is given in Appendix~\ref{app:bayes-computation}.

\section{Uncertainty summaries for experimental results}\label{app:sd}

Tables~\ref{app:two_class_sd_d1000}, \ref{app:multi_class_sd_d1000}, \ref{network_res_hamming}, and~\ref{realdata} report sample standard deviations, not standard
errors, across 50 repetitions. For gene-expression data, all methods use
the same 50 accepted splits selected by the training-only QDA-admissibility
rule in Section~\ref{sec:nutt_counts}. The means agree with the corresponding
main-text tables.

Tables~\ref{tab:s5_mcse} and~\ref{tab:proteins_mcse} separately report Monte Carlo standard errors for the S5 means in Table~\ref{1res} and the PROTEINS means in Table~\ref{tab:proteins}, respectively.

\begin{table}[H]
\centering
\footnotesize
\setlength{\tabcolsep}{3pt}
\caption{Monte Carlo standard errors for the S5 mean misclassification rates, over 50 repetitions.}
\label{tab:s5_mcse}
\begin{adjustbox}{max width=\textwidth}
\begin{tabular}{lcccccccccc}
\hline
 & Rank profile & GLDA & SVM & RF & glmnet & LightGBM & MLP & gMADD & RP-Ensemble & RDC \\
\hline
Monte Carlo SE & 0.006 & 0.008 & 0.006 & 0.007 & 0.005 & 0.006 & 0.008 & 0.005 & 0.007 & 0.006 \\
\hline
\end{tabular}
\end{adjustbox}
\end{table}

\begin{table}[H]
\centering
\footnotesize
\setlength{\tabcolsep}{2pt}
\caption{Monte Carlo standard errors for the mean balanced errors on PROTEINS, over 50 stratified splits.}
\label{tab:proteins_mcse}
\begin{adjustbox}{max width=\textwidth}
\begin{tabular}{lccccccccc}
\hline
 & RF & \shortstack{Rank\\profile} & \shortstack{RP-\\Ensemble} & WL SVM & \shortstack{gMADD-style\\WL2} & LightGBM & GLDA & glmnet & \shortstack{Radial\\SVM} \\
\hline
Monte Carlo SE & 0.002 & 0.002 & 0.002 & 0.002 & 0.002 & 0.003 & 0.002 & 0.004 & 0.003 \\
\hline
\end{tabular}
\end{adjustbox}
\end{table}

\begin{table}[H]
\centering
\footnotesize
\setlength{\tabcolsep}{6pt}
\caption{Two-class misclassification rates with standard deviations in parentheses for $d=1000$.}
\label{app:two_class_sd_d1000}
\begin{adjustbox}{max width=\textwidth}

\begin{minipage}{\textwidth}\centering
\begin{tabular}{lccccccc}
\hline
Scenario & $\mu_0$ & $a$ & Rank profile & GLDA & SVM & RF & glmnet \\
\hline
S1 & 6 & 1 & 0.023 (0.016) & 0.025 (0.017) & 0.024 (0.015) & 0.114 (0.033) & 0.200 (0.039) \\
S1 & 0 & 1.1 & 0.019 (0.016) & 0.495 (0.050) & 0.129 (0.034) & 0.421 (0.044) & 0.503 (0.037) \\
S1 & 6 & 1.1 & 0.004 (0.006) & 0.038 (0.020) & 0.005 (0.006) & 0.110 (0.035) & 0.212 (0.048) \\
S2 & 6 & 1 & 0.109 (0.034) & 0.103 (0.029) & 0.095 (0.026) & 0.159 (0.041) & 0.291 (0.065) \\
S2 & 0 & 1.1 & 0.118 (0.042) & 0.501 (0.043) & 0.189 (0.033) & 0.445 (0.044) & 0.503 (0.034) \\
S2 & 6 & 1.1 & 0.044 (0.020) & 0.127 (0.031) & 0.048 (0.020) & 0.156 (0.043) & 0.315 (0.083) \\
S3 & 6 & 1 & 0.428 (0.056) & 0.401 (0.046) & 0.396 (0.045) & 0.412 (0.056) & 0.483 (0.041) \\
S3 & 0 & 1.1 & 0.084 (0.025) & 0.501 (0.048) & 0.170 (0.037) & 0.371 (0.053) & 0.494 (0.031) \\
S3 & 6 & 1.1 & 0.084 (0.019) & 0.405 (0.050) & 0.128 (0.039) & 0.306 (0.038) & 0.492 (0.040) \\
S4 & 0 & 1 & 0.000 (0.002) & 0.468 (0.041) & 0.000 (0.000) & 0.401 (0.040) & 0.498 (0.043) \\
\hline
\end{tabular}\par\medskip
\begin{tabular}{lccccccc}
\hline
Scenario & $\mu_0$ & $a$ & LightGBM & MLP & gMADD & RP-Ensemble & RDC \\
\hline
S1 & 6 & 1 & 0.285 (0.058) & 0.170 (0.092) & 0.299 (0.047) & 0.171 (0.039) & 0.213 (0.048) \\
S1 & 0 & 1.1 & 0.496 (0.040) & 0.486 (0.049) & 0.047 (0.026) & 0.506 (0.047) & 0.591 (0.043) \\
S1 & 6 & 1.1 & 0.299 (0.070) & 0.171 (0.077) & 0.033 (0.020) & 0.189 (0.036) & 0.586 (0.044) \\
S2 & 6 & 1 & 0.327 (0.079) & 0.210 (0.076) & 0.360 (0.047) & 0.271 (0.047) & 0.272 (0.066) \\
S2 & 0 & 1.1 & 0.508 (0.038) & 0.501 (0.042) & 0.095 (0.038) & 0.498 (0.047) & 0.531 (0.030) \\
S2 & 6 & 1.1 & 0.326 (0.067) & 0.254 (0.075) & 0.067 (0.025) & 0.279 (0.047) & 0.522 (0.021) \\
S3 & 6 & 1 & 0.476 (0.052) & 0.444 (0.055) & 0.498 (0.046) & 0.449 (0.058) & 0.458 (0.042) \\
S3 & 0 & 1.1 & 0.491 (0.047) & 0.497 (0.037) & 0.109 (0.039) & 0.503 (0.057) & 0.537 (0.025) \\
S3 & 6 & 1.1 & 0.485 (0.038) & 0.452 (0.048) & 0.103 (0.035) & 0.457 (0.050) & 0.553 (0.033) \\
S4 & 0 & 1 & 0.500 (0.046) & 0.468 (0.050) & 0.007 (0.011) & 0.488 (0.049) & 0.520 (0.017) \\
\hline
\end{tabular}
\end{minipage}
\end{adjustbox}
\par\smallskip
\begin{minipage}{\textwidth}\footnotesize RDC uses half loss for exact ties, equivalent to random tie-breaking in expectation.\end{minipage}
\end{table}

\begin{table}[H]
\centering
\footnotesize
\setlength{\tabcolsep}{6pt}
\caption{Multi-class misclassification rates with standard deviations in parentheses for $d=1000$.}
\label{app:multi_class_sd_d1000}
\begin{adjustbox}{max width=\textwidth}
\begin{tabular}{lccc}
\hline
Method & S6 & S7 & S8 \\
\hline
Rank profile & 0.023 (0.012) & 0.139 (0.028) & 0.227 (0.033) \\
GLDA & 0.490 (0.036) & 0.496 (0.032) & 0.581 (0.034) \\
SVM & 0.128 (0.027) & 0.197 (0.030) & 0.310 (0.034) \\
RF & 0.453 (0.030) & 0.459 (0.035) & 0.498 (0.032) \\
gMADD & 0.053 (0.016) & 0.089 (0.020) & 0.492 (0.036) \\
glmnet & 0.517 (0.040) & 0.534 (0.035) & 0.699 (0.042) \\
LightGBM & 0.507 (0.036) & 0.510 (0.036) & 0.631 (0.051) \\
MLP & 0.481 (0.033) & 0.499 (0.030) & 0.626 (0.035) \\
RDC & 0.642 (0.042) & 0.563 (0.033) & 0.694 (0.026) \\
\hline
\end{tabular}
\end{adjustbox}
\end{table}

As a sensitivity check for gMADD, we also use the exponential
adjacency dissimilarity
\[
h(G,\tilde G)=\frac{1}{v^2}\sum_{i,j}
\left[1-\exp\{-(A_{ij}-\tilde A_{ij})^2\}\right],
\]
including the diagonal. Table~\ref{network_res_hamming} compares both inputs on
the same simulated graphs, with the gMADD classification rule unchanged.

\begin{table}[H]
\centering
\footnotesize
\setlength{\tabcolsep}{4pt}
\caption{Misclassification rates on simulated configuration-model graphs,
with standard deviations in parentheses. The first three rows
use $d_{\mathrm{net}}$ as in Table~\ref{network_res_hamming1}. The last row
uses the exponential adjacency dissimilarity $h$.}
\label{network_res_hamming}
\begin{adjustbox}{max width=\textwidth}
\begin{tabular}{lcccccc}
\hline
 & S9 ($a=4$) & S9 ($a=5$) & S9 ($a=6$) & S10 ($a=4$) & S10 ($a=5$) & S10 ($a=6$) \\
\hline
Rank profile & $\mathbf{0.278}$ (0.084) & $\mathbf{0.212}$ (0.063) & $\mathbf{0.157}$ (0.056) & $\mathbf{0.151}$ (0.058) & $\mathbf{0.104}$ (0.047) & $\mathbf{0.066}$ (0.032) \\
SVM & 0.458 (0.068) & 0.443 (0.082) & 0.449 (0.079) & 0.207 (0.063) & 0.159 (0.059) & 0.116 (0.052) \\
gMADD ($d_{\mathrm{net}}$) & 0.373 (0.084) & 0.291 (0.076) & 0.252 (0.069) & 0.268 (0.074) & 0.190 (0.071) & 0.152 (0.064) \\
gMADD ($h$) & 0.364 (0.073) & 0.285 (0.065) & 0.221 (0.071) & 0.223 (0.068) & 0.130 (0.051) & 0.091 (0.052) \\
\hline
\end{tabular}
\end{adjustbox}
\end{table}

\begin{table}[H]
\centering
\footnotesize
\setlength{\tabcolsep}{6pt}
\caption{Misclassification rates on selected real datasets with standard deviations in parentheses.}
\label{realdata}
\begin{adjustbox}{max width=\textwidth}
\begin{tabular}{lcc}
\hline
Method & Shipp & Nutt \\
\hline
Rank profile & 0.178 (0.049) & 0.168 (0.104) \\
GLDA & 0.174 (0.053) & 0.201 (0.098) \\
SVM & 0.250 (0.000) & 0.464 (0.102) \\
RF & 0.222 (0.026) & 0.249 (0.109) \\
glmnet & 0.238 (0.043) & 0.407 (0.115) \\
LightGBM & 0.250 (0.000) & 0.500 (0.000) \\
MLP & 0.168 (0.049) & 0.550 (0.080) \\
gMADD & 0.316 (0.056) & 0.369 (0.098) \\
RP-Ensemble & 0.201 (0.070) & 0.197 (0.100) \\
\hline
\end{tabular}
\end{adjustbox}
\end{table}

\section{A brief note on multilayer perceptron (MLP)}\label{appD}

    In this section, we examine the impact of different parameters in the multilayer perceptron (MLP) on high-dimensional classification tasks. All simulations are conducted under Setting ($\star$) with $F_x = N(0,1)$, $F_y = N(0,1) $, $ d = 1,000$, $\mathbf{B} = a \mathbf{A}$, $\mathbf{A} \mathbf{A}^\top = \Sigma$, and $\mu_X=0$ and $\mu_Y=\mu$, where $\mu = \mu_0  \frac{\mu'}{\Vert \mu' \Vert} $.  Here, $\Sigma_{r,c} = 0.1^{\vert r-c \vert}$, and $\mu'$ is drawn from $N_d(\mathbf{0}_d,\mathbf{I}_d)$. The test samples are generated as $  W_{x_1}, W_{x_2}, \dots, W_{x_{50}} \stackrel{\text { i.i.d }}{\sim} F_X$  and $  W_{y_1}, W_{y_2}, \dots, W_{y_{50}} \stackrel{\text { i.i.d }}{\sim} F_Y$. 

    We use an MLP structure with three hidden layers, all with the same width and activation function, followed by an output layer using the softmax activation. We first assess the impact of different activation functions in each hidden layer with $1,000$ nodes per layer and $800$ training samples.
GELU is used in the main experimental comparisons.
Table~\ref{nn_act} provides a sensitivity analysis of this choice: ReLU
and Softplus yield lower errors in the location-shift setting, whereas
GELU performs best among the tested activations in the two scale-difference
settings. All entries average 50 repetitions.

\begin{table}[H]
\centering
\footnotesize
\setlength{\tabcolsep}{6pt}
\caption{MLP mean test errors (Monte Carlo SE) with different activation functions; total training size 800.}
\label{nn_act}
\begin{adjustbox}{max width=\textwidth}
\begin{tabular}{ccccccc}
\hline
$\mu_0$ & $a$ & gelu & relu & tanh & softplus & selu \\
\hline
6 & 1 & 0.025 (0.002) & $\mathbf{0.013}$ (0.002) & 0.195 (0.005) & $\mathbf{0.006}$ (0.001) & 0.172 (0.005) \\
0 & 1.1 & $\mathbf{0.441}$ (0.007) & 0.483 (0.007) & 0.508 (0.007) & 0.491 (0.007) & 0.485 (0.008) \\
0 & 1.4 & $\mathbf{0.074}$ (0.005) & 0.489 (0.007) & 0.485 (0.008) & 0.475 (0.006) & 0.379 (0.007) \\
\hline
\end{tabular}
\end{adjustbox}
\end{table}  

Next, Table~\ref{nn_size} varies total training size, with
equal class sizes and GELU activations. Error decreases as the sample size
increases, especially for the stronger scale difference. This is a
training-size sensitivity study.
The activation and training-size grids use separate seeds, so repeated
configurations need not have identical Monte Carlo estimates.

\begin{table}[H]
\centering
\footnotesize
\setlength{\tabcolsep}{6pt}
\caption{MLP mean test errors (Monte Carlo SE) under varying total training set sizes.}
\label{nn_size}
\begin{adjustbox}{max width=\textwidth}
\begin{tabular}{cccccc}
\hline
$\mu_0$ & $a$ & 100 & 200 & 400 & 800 \\
\hline
6 & 1 & 0.176 (0.015) & 0.079 (0.007) & 0.031 (0.002) & 0.024 (0.002) \\
0 & 1.1 & 0.496 (0.007) & 0.487 (0.006) & 0.477 (0.006) & 0.455 (0.006) \\
0 & 1.4 & 0.412 (0.009) & 0.303 (0.009) & 0.133 (0.006) & 0.071 (0.005) \\
\hline
\end{tabular}
\end{adjustbox}
\end{table}  

\section{Proof of Theorem \ref{theorem1}}\label{appA}

\begingroup
Write $\Delta=\mu_Y-\mu_X$.  Independence and centering give
\begin{align*}
\mathbb E\|X_k-X_i\|_2^2
&=2\|\mathbf A\|_F^2\sigma^2(F_x),\quad k\ne i,\\
\mathbb E\|Y_k-X_i\|_2^2
&=\|\mathbf A\|_F^2\sigma^2(F_x)
 +\|\mathbf B\|_F^2\sigma^2(F_y)+\|\Delta\|_2^2.
\end{align*}
The two coordinates of $\mu_{D_X}$ follow by linearity.  Exchanging the
roles of $X$ and $Y$ gives $\mu_{D_Y}$.

It remains to verify the covariance matrices. Fix $i$.
When the required distinct indices exist, let $k_1,k_2$
be distinct indices different from $i$, and let $\ell_1,\ell_2$ be distinct $Y$-sample
indices. The corresponding identities from Definition~\ref{def1} are
\begin{align*}
\mathbb E\!\left[
 \|X_{k_1}-X_i\|_2^2\|X_{k_2}-X_i\|_2^2
\right]
&=h_1(\mathbf A,\mathbf A,\mathbf A,
\mu_X,\mu_X,\mu_X,F_x,F_x,F_x),\\
\mathbb E\|X_{k_1}-X_i\|_2^4
&=h_2(\mathbf A,\mathbf A,\mu_X,\mu_X,F_x,F_x),\\
\mathbb E\!\left[
 \|X_{k_1}-X_i\|_2^2\|Y_{\ell_1}-X_i\|_2^2
\right]
&=h_1(\mathbf A,\mathbf A,\mathbf B,
\mu_X,\mu_X,\mu_Y,F_x,F_x,F_y),\\
\mathbb E\!\left[
 \|Y_{\ell_1}-X_i\|_2^2\|Y_{\ell_2}-X_i\|_2^2
\right]
&=h_1(\mathbf A,\mathbf B,\mathbf B,
\mu_X,\mu_Y,\mu_Y,F_x,F_y,F_y),\\
\mathbb E\|Y_{\ell_1}-X_i\|_2^4
&=h_2(\mathbf A,\mathbf B,\mu_X,\mu_Y,F_x,F_y).
\end{align*}
Expanding the square of each sample average and separating equal from unequal
indices therefore yields
{\allowdisplaybreaks[0]
\begin{align*}
\operatorname{Var}\{D_X(X_i)\}
&=\frac{n-2}{n-1}h_{1,XXX}
 +\frac{1}{n-1}h_{2,XX}-\{\mathbb ED_X(X_i)\}^2,\\
\operatorname{Var}\{D_Y(X_i)\}
&=\frac{m-1}{m}h_{1,XYY}
 +\frac{1}{m}h_{2,XY}-\{\mathbb ED_Y(X_i)\}^2,\\
\operatorname{Cov}\{D_X(X_i),D_Y(X_i)\}
&=h_{1,XXY}-\mathbb ED_X(X_i)\,\mathbb ED_Y(X_i),
\end{align*}
}
where the abbreviated $h$ terms are the five quantities displayed above.
When $n=2$, the within-$X$ off-diagonal sum is empty,
and its coefficient $(n-2)/(n-1)$ is zero. The analogous within-$Y$
off-diagonal sum is empty when $m=2$. Thus the covariance formulas remain
valid without requiring nonexistent sample indices in these boundary cases.
Substitution of the two means gives exactly the entries of $\Sigma_{D_X}$ in
Definition~\ref{def2}.  The entries of $\Sigma_{D_Y}$ follow by exchanging
$(\mathbf A,\mu_X,F_x,n)$ with $(\mathbf B,\mu_Y,F_y,m)$ and replacing
$\Delta$ by $-\Delta$.  This proves the theorem.

For an independent test observation, the same calculation includes all $n$
or $m$ training distances.  The equal-index weights consequently become
$1/n$ and $1/m$, respectively, which gives the covariance matrices in
Theorem~\ref{theorem4}.
\endgroup

\section{Proof of the Gaussian approximation}\label{appB}

A useful specialization of (C1)--(C2) is the banded
Toeplitz-type case in which the components of $\Delta$ are equal and the
nonzero diagonals of $\mathbf A$ and $\mathbf B$ are constant.  Conditions
(C3)--(C4) additionally require Lyapunov decay and exclude asymptotic
degeneracy of the two-dimensional profile.
The fourth-moment assumption in Setting~$(\star)$ suffices for the exact
moment formulas and the remainder bound below; it does not imply (C3).
Condition (C3) separately requires third absolute moments of quadratic block
sums. For example, with $\mathbf A=\mathbf B=I_d$ and $t_5$ latent
coordinates, these moments are infinite, so this CLT does not cover that
case, although the exact mean and covariance formulas still apply.

\paragraph{Proof of Theorem~\ref{thm:distance_clt}.}

By exchangeability within each class, it suffices to prove the result for
$D(X_1)$; the argument for $D(Y_j)$ is identical.  Write
\begin{align*}
D(X_1)-\mathbb E\{D(X_1)\}
&=\sum_{k=1}^d V_k^{(X)},\\*
V_k^{(X)}
&=\begin{pmatrix}
\displaystyle \frac{1}{n-1}\sum_{\substack{1\le\ell\le n\\\ell\ne1}}
\left[(\mathbf a_k^\top x_\ell-\mathbf a_k^\top x_1)^2
-\mathbb E[(\mathbf a_k^\top x_\ell-\mathbf a_k^\top x_1)^2]\right]\\[2mm]
\displaystyle \frac{1}{m}\sum_{\ell=1}^m
\left[(\mathbf b_k^\top y_\ell-\mathbf a_k^\top x_1+\Delta_k)^2
-\mathbb E[(\mathbf b_k^\top y_\ell-\mathbf a_k^\top x_1+\Delta_k)^2]\right]
\end{pmatrix}.
\end{align*}
Condition~(C1) implies that $V_k^{(X)}$ and $V_{k'}^{(X)}$ are independent
whenever $|k-k'|>2m_0$.  Because the definition of $R_q^{(X)}$ leaves a
$2m_0$-coordinate boundary at the end of each block, the vectors
$R_1^{(X)},\ldots,R_t^{(X)}$ are independent.  Let
\[
S_d^{(X)}=\sum_{q=1}^tR_q^{(X)},\qquad
T_d^{(X)}=D(X_1)-\mathbb E\{D(X_1)\}-S_d^{(X)}.
\]
By definition, $\operatorname{Cov}(S_d^{(X)})=\Sigma_t^{(X)}$.  Hence the
summands $(\Sigma_t^{(X)})^{-1/2}R_q^{(X)}$ are centered, independent, and
have covariance matrices summing to $I_2$.  Condition~(C3) and the multivariate
Berry--Esseen bound of \mbox{\citet{raivc2019multivariate}} therefore give
\[
(\Sigma_t^{(X)})^{-1/2}S_d^{(X)}\xrightarrow{d}N_2(0,I_2).
\]

It remains to control the omitted block boundaries and the final $r$
coordinates. Conditions~(C1)--(C2) and the finite fourth moments in
Setting~$(\star)$ imply a bound on $\mathbb E\|V_k^{(X)}\|_2^2$
uniform in both $k$ and $d$.
Finite-range dependence then yields
\[
\mathbb E\|T_d^{(X)}\|_2^2
=O\{t(2m_0)^2+r^2\}
=O(d^{1-\alpha})+O(d^{2\alpha})=o(d),
\]
because $\alpha<1/2$.  By Condition~(C4),
\[
\mathbb E\left\|(\Sigma_t^{(X)})^{-1/2}T_d^{(X)}\right\|_2^2
\leq\frac{1}{cd}\mathbb E\|T_d^{(X)}\|_2^2=o(1).
\]
Thus $(\Sigma_t^{(X)})^{-1/2}T_d^{(X)}=o_p(1)$, and Slutsky's theorem proves
the covariance-standardized conclusion.  If
$d^{-1}\Sigma_t^{(X)}\to V_X\succ0$, the continuous mapping theorem gives
the stated $d^{-1/2}$-scaled limit.  The proof for $D(Y_j)$ is the same.
For an independent test profile, replacing $x_1$ or $y_1$ by the corresponding
independent test vector produces the analogous coordinate contributions and
block decomposition, so the same argument applies under the test-profile
versions of Conditions~(C3)--(C4).

Finally, the full profile covariance and the block covariance are
asymptotically equivalent in the normalization used above. Indeed, put
$U_d=(\Sigma_t^{(X)})^{-1/2}S_d^{(X)}$ and
$E_d=(\Sigma_t^{(X)})^{-1/2}T_d^{(X)}$. Then
$\operatorname{Cov}(U_d)=I_2$ and $\mathbb E\|E_d\|_2^2=o(1)$.
Cauchy--Schwarz bounds each entry of $\operatorname{Cov}(U_d,E_d)$ by
$o(1)$, whence
\[
(\Sigma_t^{(X)})^{-1/2}\Sigma_{D_X}(\Sigma_t^{(X)})^{-1/2}
=I_2+o(1).
\]
The same argument applies to $Y$ and to independent test profiles, with
$\Sigma_{D_{W_g}}$ replacing $\Sigma_{D_g}$. This justifies using the full
test-profile covariance in the Gaussian reference.

\end{document}